\documentclass[a4paper,11pt]{article}
\pdfoutput=1 

\usepackage[T1]{fontenc} 
\usepackage[utf8]{inputenc}
\usepackage{microtype}
\usepackage{lmodern}
\usepackage[english]{babel}
\usepackage{siunitx}
\usepackage{xspace}
\usepackage{jcappub} 

\usepackage[export]{adjustbox}  

\title{\boldmath Signal model and event reconstruction for the radio detection of inclined air showers}

\author[a,b,1]{F. Schlüter\begin{NoHyper}\note{Corresponding author.}\end{NoHyper}} 
\author[a,c]{and T. Huege}

\affiliation[a]{Karlsruher Institut für Technologie, Institut für Astroteilchenphysik, Karlsruhe, Germany}
\affiliation[b]{Universidad Nacional de San Martín, Instituto de Tecnologías en Detección y Astropartículas,\\ Buenos Aires, Argentina}
\affiliation[c]{Vrije Universiteit Brussel, Astrophysical Institute, Brussels, Belgium}
\emailAdd{fschluter@icecube.wisc.edu}
\emailAdd{tim.huege@kit.edu}

\abstract{
    The detection of inclined air showers (zenith angles $\theta \gtrsim 65^\circ$) with kilometer-spaced radio-antenna arrays allows measuring cosmic rays at ultra-high energies ($E \lesssim 10^{20}\,\mathrm{eV}$). Radio and particle detector arrays provide independent measurements of the electromagnetic and muonic shower components of inclined air showers, respectively. Combined, these measurements have a large sensitivity to discriminate between air showers initiated by lighter and heavier cosmic rays.

    We have developed a precise model of the two-dimensional, highly complex and asymmetric lateral radio-signal distributions of inclined air shower at ground -- the ``radio-emission footprints''. Our model explicitly describes the dominant geomagnetic emission with a rotationally symmetric lateral distribution function, on top of which additional effects disturb the symmetry. The asymmetries are associated with the interference between the geomagnetic and sub-dominant charge-excess emission as well as with geometrical projection effects, so-called ``early-late'' effects. Our fully analytic model describes the entire footprint with only two observables: the geometrical distance between the shower impact point at the ground and the shower maximum $d_\mathrm{max}$, and the geomagnetic radiation energy $E_\mathrm{geo}$. We demonstrate that with this model, the electromagnetic shower energy can be reconstructed by kilometer-spaced antenna arrays with an intrinsic resolution of 5\% and a negligible bias. 

}

\newcommand{\eV}[1]{\ensuremath{10^{#1}\,\text{eV}}\xspace}
\newcommand{\gcm}{\ensuremath{\text{g}\,\text{cm}^{-2}}\xspace}
\newcommand{\kgm}{\ensuremath{\text{kg}\,\text{m}^{-3}}\xspace}

\newcommand{\Eem}{\ensuremath{E_\mathrm{em}}\xspace}
\newcommand{\EemMC}{\ensuremath{E_\mathrm{em}^\mathrm{MC}}\xspace}
\newcommand{\Egeo}{\ensuremath{E_\mathrm{geo}}\xspace}
\newcommand{\Sgeo}{\ensuremath{S_\mathrm{geo}}\xspace}
\newcommand{\dmax}{\ensuremath{d_\mathrm{max}}\xspace}
\newcommand{\Xmax}{\ensuremath{X_\mathrm{max}}\xspace}

\newcommand{\rhomax}{\ensuremath{\rho_\mathrm{max}}\xspace}

\newcommand{\fgeo}{\ensuremath{f_\mathrm{geo}}\xspace}
\newcommand{\fce}{\ensuremath{f_\mathrm{ce}}\xspace}
\newcommand{\fgs}{\ensuremath{f_\mathrm{GS}}\xspace}
\newcommand{\ace}{\ensuremath{a_\mathrm{ce}}\xspace}

\newcommand{\vB}{\ensuremath{\vec{v} \times \vec{B}}\xspace}
\newcommand{\vxB}{\ensuremath{\vec{v} \times \vec{B}}\xspace}

\newcommand{\vvB}{\ensuremath{\vec{v} \times  (\vec{v} \times \vec{B})}\xspace}
\newcommand{\vxvxB}{\ensuremath{\vec{v} \times  (\vec{v} \times \vec{B})}\xspace}

\newcommand{\fvB}{\ensuremath{f_{\vec{v} \times \vec{B}}}\xspace}
\newcommand{\fvxB}{\ensuremath{f_{\vec{v} \times \vec{B}}}\xspace}
\newcommand{\fvvB}{\ensuremath{f_{\vec{v} \times  (\vec{v} \times \vec{B})}}\xspace}
\newcommand{\fv}{\ensuremath{f_{\vec{v}}}\xspace}

\makeatletter
\gdef\@fpheader{Published as \href{https://iopscience.iop.org/article/10.1088/1748-0221/14/04/P04005}{\textit{JINST} \textbf{14} (2019) P04005}}
\makeatother

\begin{document}
\maketitle
\flushbottom


\section{Detection of inclined air showers with radio antennas}
The detection of inclined air showers with radio antennas has recently been demonstrated by the Pierre Auger Observatory \cite{Aab:2018ytv}. The radio emission from those showers can illuminate large footprints of several square kilometers, as had been previously predicted via simulations \cite{Huege:2015lga}, eventually exceeding the footprints measurable by particle detectors \cite{Gottowik:2021tds}. This enables the detection of ultra-high energy cosmic rays (UHECRs) up to the highest energies, e.g., \eV{20}, with radio antennas, as antennas can be sparsely-spaced such that the instrumented area provides sufficient aperture ($>$\,\SI{1000}{km^2\,sr}). The Radio Detector of the upgraded Pierre Auger Observatory \cite{2019ICRC_pont}, which will consist of over 1600 radio-antenna stations covering an area of nearly \SI{3000}{km^2}, will routinely detect such events. Other experiments such as the envisioned Giant Radio Array for Neutrino Detection (GRAND) aim to detect inclined air showers from UHECRs as well as from ultra-high-energy neutrinos \cite{Alvarez-Muniz:2018bhp}.

Radio antennas measure the electromagnetic radiation emitted by electrons and positrons during their propagation through the atmosphere. While most electrons and positrons are ultimately absorbed in the atmosphere, the radio emission experiences no significant attenuation and can be measured by radio antennas tens to hundreds of kilometers away from the emission region (here approximated as the location of the shower maximum). At such distances, particle detectors mostly measure muons while other particles are absorbed in the atmosphere. Measurements of the showers' muon content suffer from a significant ambiguity between the primary cosmic-ray energy and mass, whereas the strength of the electromagnetic radiation has no significant correlation with the primary mass. Combining both measurements allows us to infer the cosmic-ray mass and test hadronic interaction models \cite{PierreAuger:2014ucz,PierreAuger:2016nfk,Holt:2019fnj,PierreAuger:2021qsd}.

For such studies, an accurate reconstruction of the shower energy from radio measurements is indispensable. For vertical showers, i.e., showers with zenith angles $\theta < \SI{60}{\degree}$, several signal models of the radio emission at ground to reconstruct the shower energy have been proposed \cite{Nelles:2014xaa,Glaser:2018byo} and used with experimental data \cite{PierreAuger:2015hbf,PierreAuger:2016vya}. In those models, the radio-emission footprints are described based on the macroscopic interpretation of the superposition of two emission mechanisms, the charge-excess (Askaryan) and geomagnetic emission \cite{Werner:2007kh,deVries:2010ti}. Additionally, the models account for the temporal ``Cherenkov'' compression of the radio emission, where at a characteristics distance around the shower axis, the emission from the entire longitudinal development arrives almost simultaneously causing an enhancement of the coherent signal \cite{deVries:2011pa,Alvarez-Muniz:2012ytl}. This imprints an annulus in the emission pattern, here referred to as Cherenkov ring. 

Those models are not applicable for inclined air showers, though, as both the interference between the emission mechanisms and the Cherenkov ring are known to change with the ambient atmospheric conditions in the emission region of the air shower and hence with the zenith angle. Furthermore, for inclined air showers, the radio emission is strongly ``projected'' onto the ground plane. This projection imprints geometrical early-late effects which disturbs the interference pattern between the both aforementioned mechanisms and becomes significant for zenith angles beyond \SI{60}{\degree} \cite{Huege:2018kvt}. 

Therefore, we present a model dedicated to the description of the radio-emission footprints from inclined air showers. The lateral distribution of the geomagnetic emission is individually described by a 1-dimensional lateral distribution function (LDF) after the emission has been corrected for the aforementioned early-late ``asymmetry''. With increasing zenith angle, the relative strength of the charge-excess emission decreases and the geomagnetic emission dominates the total signal. Hence, the interference between both emission mechanisms is treated as a further asymmetry to the geomagnetic emission. Using a comprehensive set of CoREAS simulations we found that the shape of the LDF for the geomagnetic emission as well as the asymmetries can be described with a single parameter, \dmax, the geometrical distance between shower impact position at the ground,
in the following also refereed to as ``shower core'', and the shower maximum. The position of the shower maximum, and hence \dmax, scales in first order with the zenith angle and in second order with the depth of the shower maximum \Xmax. Lastly, the amplitude of the geomagnetic emission can be described by the geomagnetic radiation energy \Egeo, i.e., the spatial integral over the energy deposit of the geomagnetic emission at ground. Hence, the entire emission footprint is described by two observables: \Egeo and \dmax.


Our model describes the radio emission in terms of energy fluence $f$ [eV$\,$m$^{-2}$], i.e., the energy deposit per unit area, in the \SIrange{30}{80}{MHz} band. This frequency band is used by most current-generation large-scale radio detector arrays \cite{Holt:2017dyo,Schellart:2013bba,Bezyazeekov:2015rpa} and in particular by the Auger Radio Detector. The simulations used here describe the conditions of the Pierre Auger Observatory \cite{2015172}, located near the city of Malargüe, Argentina, in the Southern Hemisphere. This concerns the local magnetic field (strength and orientation), the observation height (altitude), and atmospheric conditions. The adaptability of this model to other conditions, including other frequency bands as many next-generation radio experiments \cite{Huege:2016jvc,Alvarez-Muniz:2018bhp,Schroder:2019suq} aim to cover higher frequencies and larger bands, is discussed in Sec. \ref{sec:ldf:discussion}.

This article is structured as follows: In section \ref{sec:ldf:data}, we discuss the simulation sets utilized to derive the radio-emission footprint model and reconstruction algorithm as well as its evaluation. A qualitative description of the radio-emission footprints and their asymmetries is given in section \ref{sec:ldf:emission}. In section \ref{sec:ldf:model}, we develop the signal model. In section \ref{sec:ldf:rec_geomagnetic_energy}, the electromagnetic shower energy is reconstructed, and the intrinsic performance is evaluated with simulations of a kilometer-sparse antenna array. Finally, we discuss and conclude in sections \ref{sec:ldf:discussion} and \ref{sec:ldf:conclusiuon}.
\section{Simulation and signal processing of the radio-emission footprints of inclined air showers}
\label{sec:ldf:data}
We use two different sets of air shower simulations, one to develop the model for the radio-emission footprints, and one to evaluate the reconstruction of the electromagnetic shower energy with this model. The sets differ mainly in their detector layout, i.e., in the positioning of the observers, and the coverage of the phase space, i.e., the distributions for the showers' energy and arrival direction. For development, we use simulations with an artificial, unrealistically dense detector layout with an antenna grid which is centered around the shower core. The phase space is covered uniformly by discrete, equidistant bins in energy, zenith angle, and azimuth angle. For validation, we use simulations with a realistic, sparse detector layout with showers randomly located within a finite array and a phase space that is sampled continuously in energy and arrival direction.

For all air showers, the particle cascades are simulated with CORSIKA \cite{corsika} and the radio emission is calculated with the CoREAS extension \cite{Huege:2013vt}. The detector arrays are placed at an altitude of \SI{1400}{m} a.s.l.\ and in a local magnetic field matching the conditions at the site of the Pierre Auger Observatory in Argentina with an inclination of $\sim\,$\SI{-36}{\degree} and a strength of $\sim\,$\SI{0.24}{G}. If not mentioned otherwise, the simulated atmosphere, i.e., the density profile $\rho(h)$ as function of the altitude $h$ and refractive index at sea level of $n = 1 + 3.12 \cdot 10^{-4}$ are used to match the conditions at the Pierre Auger Observatory in October. For the simulation of the particle cascades, an electron multiple-scattering-length factor ``STEPFC'' of 1 was used. It has been reported previously that lowering this parameter to 0.05 increases the total emitted radiation energy by 11\% regardless of the zenith angle and energy \cite{Gottowik:2017wio}, but increases the computational effort per shower by a factor of $\sim 4$. Hence, we choose to retain a value of 1 with the consequence that the final normalization of our model has to be adjusted for the missing 11\% of radiation energy. Below, details to the detector layout for the two simulations sets are given, additional information is summarized in Table \ref{tab:simulation:settings}.

For development, we have simulated 4309 showers in which the radio-emission footprint is sampled at 240 observers situated on a flat ground plane such that a star-shaped grid with 8 rays and equidistant antenna spacing is formed in a shower-plane coordinate system perpendicular to the air shower arrival direction (cf. Fig. \ref{fig:ldf:shower_ground_plane}, both panels). Within the shower-plane coordinates, the observers are placed depending on the orientation w.r.t.\ the magnetic field vector to allow for an optimal decomposition of the geomagnetic and charge-excess emission, as it will become clear later.

For evaluation, we have simulated 15970 air showers for which the shower core is randomly distributed within a finite hexagonal array with a spacing of \SI{1500}{m}. The array resembles that of the Pierre Auger Observatory and extends across nearly \SI{3000}{km^2}. For each shower, all observers within a zenith-angle-dependent maximum distance to the shower axis are simulated. 

\begin{table}[t]
    \centering
    \caption{CORSIKA simulation settings for the two different simulation sets.}
    \vspace{0.2cm}
    \begin{tabular}{c|c|c}
         & Development & Validation \\
         \hline
         Release & v7.69\footnotemark & v7.7 \\ 
         Number of showers & 4309 & 15970 \\
         Primaries & p, Fe & p, He, N, Fe \\
         Energies $E$ / eV & $10^{18.4}$, $10^{18.6}$, .. $10^{20.2}$ & [$10^{18.4}$, $10^{20.1}$] flat in $\log_{10}$ \\
         Zenith angles $\theta$ & 65$^\circ$, 67.5$^\circ$, .. 85$^\circ$ & [65$^\circ$, 85$^\circ$] flat in $\sin^2$ \\
         Azimuth angles $\phi$ & 0$^\circ$, 45$^\circ$, .. 315$^\circ$ & [0$^\circ$, 360$^\circ$) \\
         h.e. had. int. model & QGSJETII-04 \cite{Ostapchenko:2010vb} & QGSJETII-04, Sibyll2.3d \cite{Ahn:2009wx,Riehn:2019jet}\\
         l.e. had. int. model & UrQMD \cite{Bleicher:1999xi} & UrQMD \\
         Thinning $\epsilon_\mathrm{thin}$ & $5 \times 10^{-6}$ & $1 \times 10^{-6}$ \\
    \end{tabular}
    \label{tab:simulation:settings}
\end{table}
\footnotetext{This version was modified with an optimization for inclined air showers which was published with v7.7.}    

In addition to the 4309 simulations with the star-shaped antenna grid, we simulated three times 216 proton showers with an energy of $\log_{10}{(E / \si{eV})} = 18.4$ and varying atmospheric conditions. These simulations cover the same arrival directions, have 160 observers on a star-shaped grid and a refined particle thinning of $\epsilon_\mathrm{thin} = 1 \times 10^{-6}$. The atmospheric conditions match those at the site of the Pierre Auger Observatory in February and June, as well as the US standard atmosphere, as provided within CORSIKA. 

In the following, we refer to the atmospheres also with their CORSIKA IDs: US standard: 1, February: 19, June: 23, October: 27. For developing the signal model and reconstruction of air showers, we rely on a model of the atmosphere, i.e., a model for the atmospheric density gradient, from \cite{PhDGlaser,radiotools}, which was extended and improved in the context of this work to replicate the atmosphere simulated in CORSIKA/CoREAS.

We use thinning to reduce the considerable computational effort. However, thinning affects the simulation of weak radio signals at large axis distances which have to be treated with caution. This is explained and addressed in more detail in Appendix \ref{app:ldf:thinning}. 

CORSIKA computes a Gaisser-Hillas fit to the energy-deposit table to determine the depth of the air-shower maximum, \Xmax. We found that this fit does not reliably work for air showers with zenith angles beyond \SI{80}{\degree}. Hence, we perform a simple 2-step $\chi^2$ minimization ourselves to determine the depth of shower maximum. The resulting \Xmax distribution is uncorrelated with the zenith angle \cite[Figure 4.1]{Schluter:2022yev}. 

The electric field pulses are simulated in the North-South (NS), West-East (WE), and Vertical (V) polarizations. From the time series of each polarization, the energy fluence is calculated by a sum over the squares of the electric field amplitudes in a \SI{100}{ns} time interval centered around the peak \cite[Eq. 1]{PierreAuger:2015hbf}\footnote{For simulations without noise the second term in the equation is not necessary.}. The peak is defined as the maximum of the quadratic sum of the Hilbert envelopes from all 3 polarizations. The simulated pulses are band-pass filtered to the \SIrange{30}{80}{MHz} band with an idealized rectangle filter. A frequency resolution of $\sim 100\,$kHz is ensured by zero-padding the traces sufficiently. For developing our signal model we have to decompose the radio emission at an observer in parts originating from the geomagnetic and charge-excess emission. This is accomplished exploiting the known polarization characteristics of both emission mechanisms and is explained in Appendix \ref{app:ldf:decomposition}.

\section{Radio emission from inclined air showers}
\label{sec:ldf:emission}
\begin{figure}[t]
    \center
    \includegraphics[width=\textwidth]{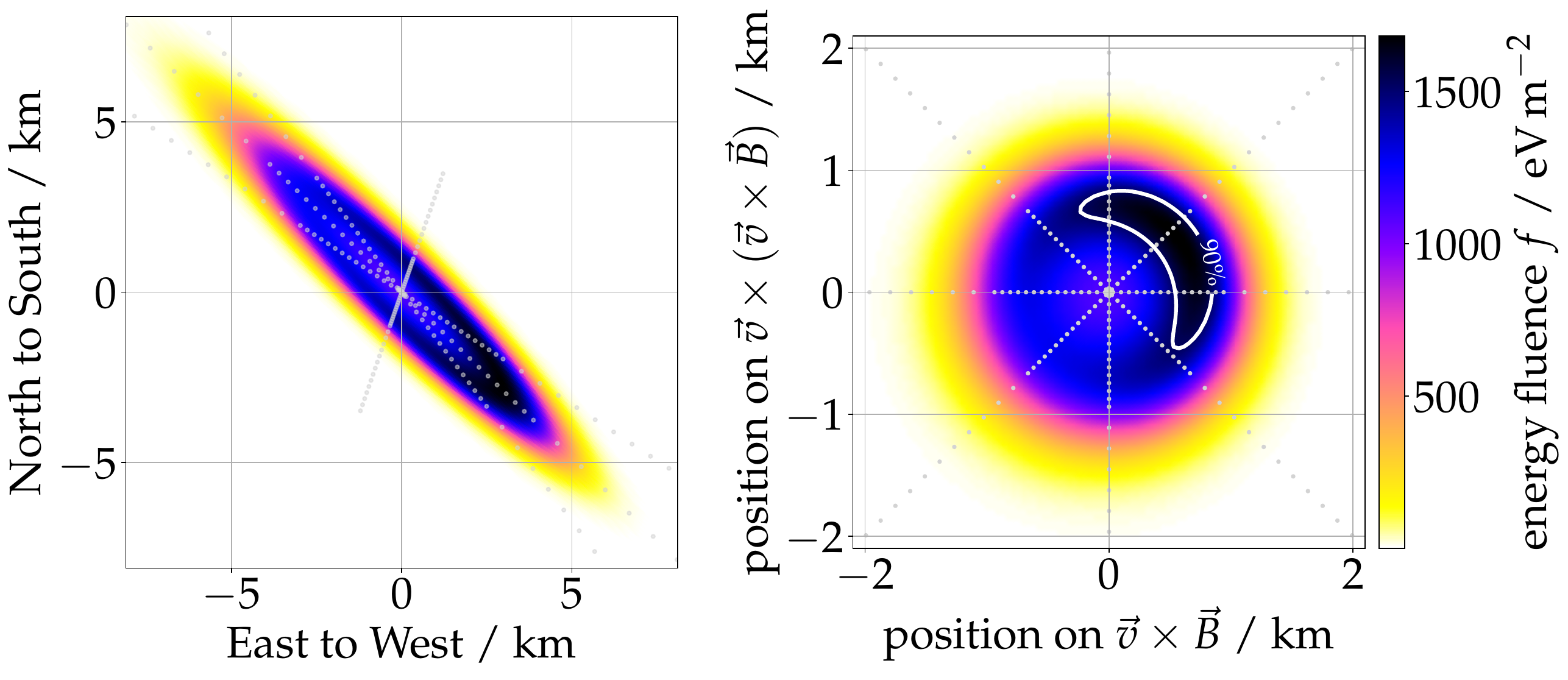}
    \caption{The simulated radio emission of an 80$^\circ$ CoREAS shower arriving from South-East is shown in energy fluence (color coded) for \SIrange{30}{80}{MHz}. The emission is simulated for 240 observers (indicated by the gray dots) situated at ground and interpolated in between them using Fourier modes \cite{Fourier}. \textit{Left}: Largely elongated and strongly asymmetric radio-emission footprint in ground-plane coordinates with the shower incoming from the bottom-right. \textit{Right}: The same footprint in the commonly used \vxB, \vxvxB shower-plane coordinates. The white band indicates the contour of 90\% the maximum signal. The emission pattern is inconsistent with the typical interference pattern of the both emission mechanisms as it is not symmetric w.r.t. the \vxB axis, see details in text.}
    \label{fig:ldf:shower_ground_plane}
\end{figure}
Figure \ref{fig:ldf:shower_ground_plane} shows the radio-emission footprint at ground from a (simulated) \SI{80}{\degree} air shower coming from South-East. The color map shows the energy fluence $f$. The panel on the left shows the footprint on the ground plane which is highly elongated along the shower direction covering a large area with a semi-major axis of $\sim\,$\SI{10}{km} and a semi-minor axis of $\sim\,$\SI{2}{km}, and exhibits strong asymmetries. The right panel shows the radio-emission footprint in a shower-plane coordinate system. In this representation, the footprint is more circular. The coordinate system is defined by orthogonal unit vectors pointing into the directions \vB, \vvB, and $\vec{v}$, where $\vec{v}$ is the direction of the primary particle trajectory (i.e., pointing exactly in the opposite direction of the shower axis), and $\vec{B}$ pointing in the direction of the magnetic field vector which points upwards at the location (latitude) of the Pierre Auger Observatory. This coordinate system is commonly used to display the radio emission from extensive air showers as it highlights the interference between the geomagnetic and charge-excess emission: along the \vxB axis the interference is maximal while both emission mechanisms are disentangled along the \vxvxB axis; see \cite{Huege:2016veh} for a comprehensive review of the emission pattern of the radio emission.

The footprint in the shower-plane coordinates shows a strong asymmetry (roughly) along the \vxB axis (x-axis) which is known to originate from the superposition of geomagnetic and charge-excess emission. However, the footprint is not symmetric w.r.t.\ the \vxB axis as highlighted by the white contour marking 90\% of the maximum fluence which is found to be rotated counter-clockwise w.r.t.\ the \vxB axis by 38$^\circ$. This inconsistency from the interference pattern of the two emission mechanisms can be explained with the so-called early-late asymmetry. For non-vertical showers, observers at the ground which are below the shower axis will measure the radio emission at an ``earlier'' stage of the shower development, i.e., they are closer to the point of emission (in this work assumed to be at \Xmax), than observers above the shower axis. Therefore, the expanding electric field will have a higher intensity and consequently, an early observer will measure a stronger signal than a late observer\footnote{This assumes that the showers are fully developed, i.e., the electromagnetic component responsible for the radio emission is already absorbed, which is given for inclined air showers.}. Additionally, an early and late observer with equal distances to the shower axis will not have the same off-axis or viewing angle, i.e., not the same angle between the line-of-sight from \Xmax to the observer and the shower axis. An illustration of the differences between an early and late observer is given in Fig.~\ref{fig:ldf:early-late-diagram}. Both effects will introduce an asymmetry in the lateral distribution of the emission which becomes relevant only beyond a zenith angle of \SI{60}{\degree} and increases with the distance to the shower maximum \dmax and the axis distance of the observer. Correcting for these effects reduces the asymmetry in our simulated radio-emission footprints and restores the known asymmetry pattern from the interference of the geomagnetic and charge-excess emission, as we will see later. A qualitative analysis of the position of the maximum energy fluence in shower-plane coordinates before and after correcting for the early-late asymmetry can be found in Appendix \ref{app:ldf:el-angle}. Finally, when subtracting the charge-excess emission for the overall emission, we are left with the rotational symmetric geomagnetic emission. 

While the overall asymmetry in the radio-emission footprints is dominated by the interference between the two emission mechanisms in the lower half of the zenith-angle range we consider here, the early-late effects constitute the dominant asymmetry for the upper half of the zenith-angle range. Note, that this changes for experiments located at different locations on Earth depending on the strength of the local geomagnetic field.

Besides the asymmetry, the ring-like structure of the temporal Cherenkov compression, i.e., the Cherenkov ring, is visible in the emission pattern. The radius of this ring, i.e., the Cherenkov radius $r_0$, can be estimated from the base of a cone with its apex at the shower maximum with an opening angle equal to the Cherenkov angle $\delta_\mathrm{Che}(h = h_\mathrm{max})$. For a point source that is moving with the speed of light $\beta = 1$, the radius is
\begin{equation}
    \label{eq:ldf:r0}
    r_0 = \tan(\delta_\mathrm{Che}) \, \dmax \hspace{0.2cm} \text{with} \hspace{0.2cm} \delta_\mathrm{Che} = \cos^{-1}\left(1 / n(h = h_\mathrm{max})\right),
\end{equation}
where $n(h_\mathrm{max})$ is the refractive index at the shower maximum which is a function of the altitude or height above sea level $h$.

Recently, an additional ``apparent'' asymmetry in the radio-emission footprint of very inclined air showers with zenith angle beyond 80$^\circ$ has been reported \cite{Gottowik:2021tds}. In \cite{Schluter:2020tdz} it is shown that this apparent asymmetry can be explained and resolved by a displacement of the whole radio-emission footprint w.r.t.\ the Monte-Carlo (MC) shower core. This core displacement is explained by the refraction of the radio emission during propagation in the Earth's atmosphere. Here, we account for it by allowing the core coordinates, i.e., the coordinates of the radio symmetry center, to vary from the MC core. The coordinates are found fitting the lateral signal distribution, cf.\ Sec.~\ref{sec:ldf:geomagnetic_emission}. The displacement also implies that \dmax changes if calculated between the shower maximum and the displaced core instead of the MC core. However, the effect to \dmax is below 1\% for all zenith angles and thus ignored in the following. 

\section{Model for the radio-emission footprints}
\label{sec:ldf:model}
To describe the radio-emission footprint from inclined air showers, we first have to remove the early-late asymmetry. In Sec.~\ref{sec:ldf:early-late}, a purely geometrical correction for this asymmetry is formulated and evaluated. With this asymmetry removed, we can determine a parameterization of the shape of the symmetric geomagnetic emission in Sec.~\ref{sec:ldf:geomagnetic_emission}. Our approach to subtract the charge-excess from the geomagnetic emission and describe the interference between the two emission mechanisms, respectively, is described in Sec.~\ref{sec:ldf:charge_excess_fraction}. In Sec.~\ref{sec:ldf:rec_geomagnetic_energy}, the geomagnetic radiation energy \Egeo is introduced.

\subsection{Geometrical early-late effects}
\label{sec:ldf:early-late}
\begin{figure}[t]
    \center
    \includegraphics[width=.95\textwidth]{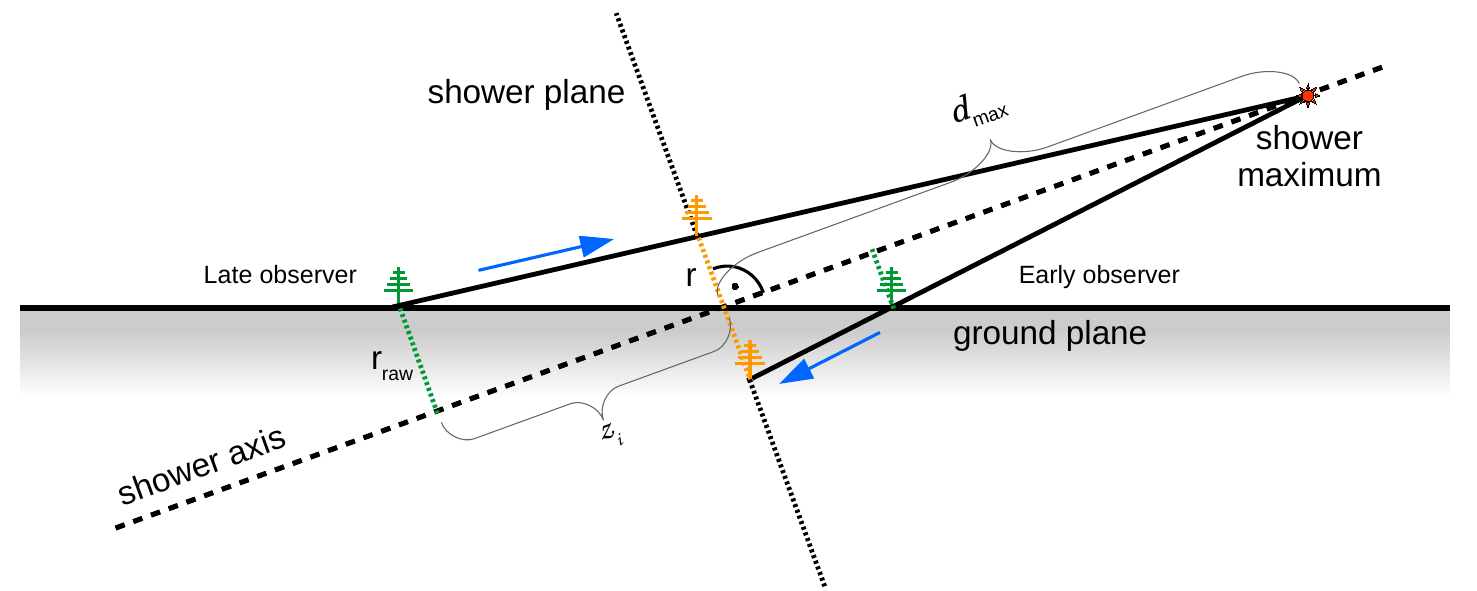}
    \caption{Illustration of an inclined air shower with an early and late observer at the same early-late corrected axis distance, i.e., with the same off-axis angle. To correct for the early-late effects which cause the asymmetry, we ``project'' signals measured at ground (indicated with green antenna symbol) along the line of sight from antenna to shower maximum into the shower plane (orange antennas). See text for details.}
    \label{fig:ldf:early-late-diagram}
\end{figure}
\begin{figure}[t]
    \centering
    \includegraphics[valign=t,width=.46\textwidth]{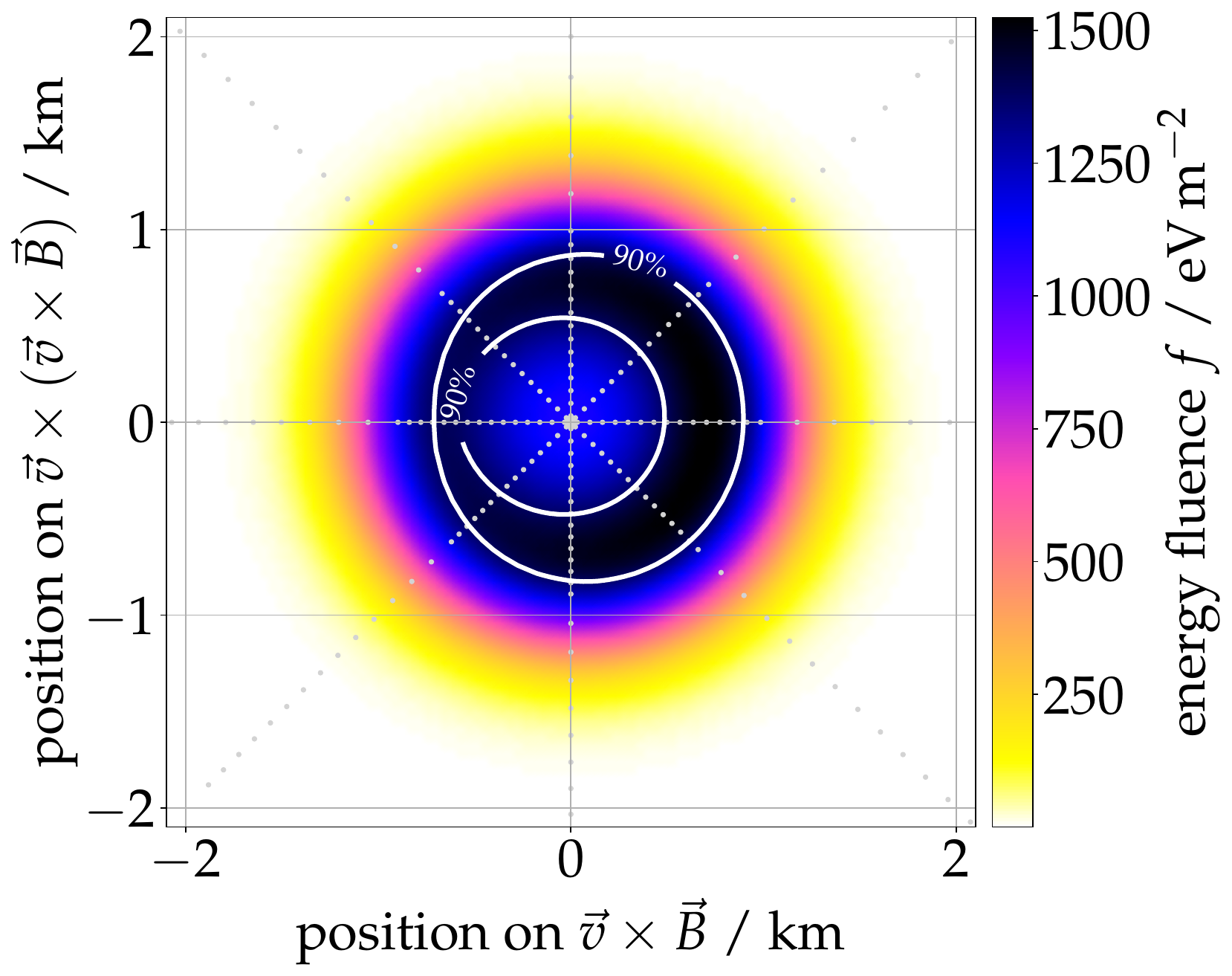}\hfill
    \includegraphics[valign=t,width=.51\textwidth]{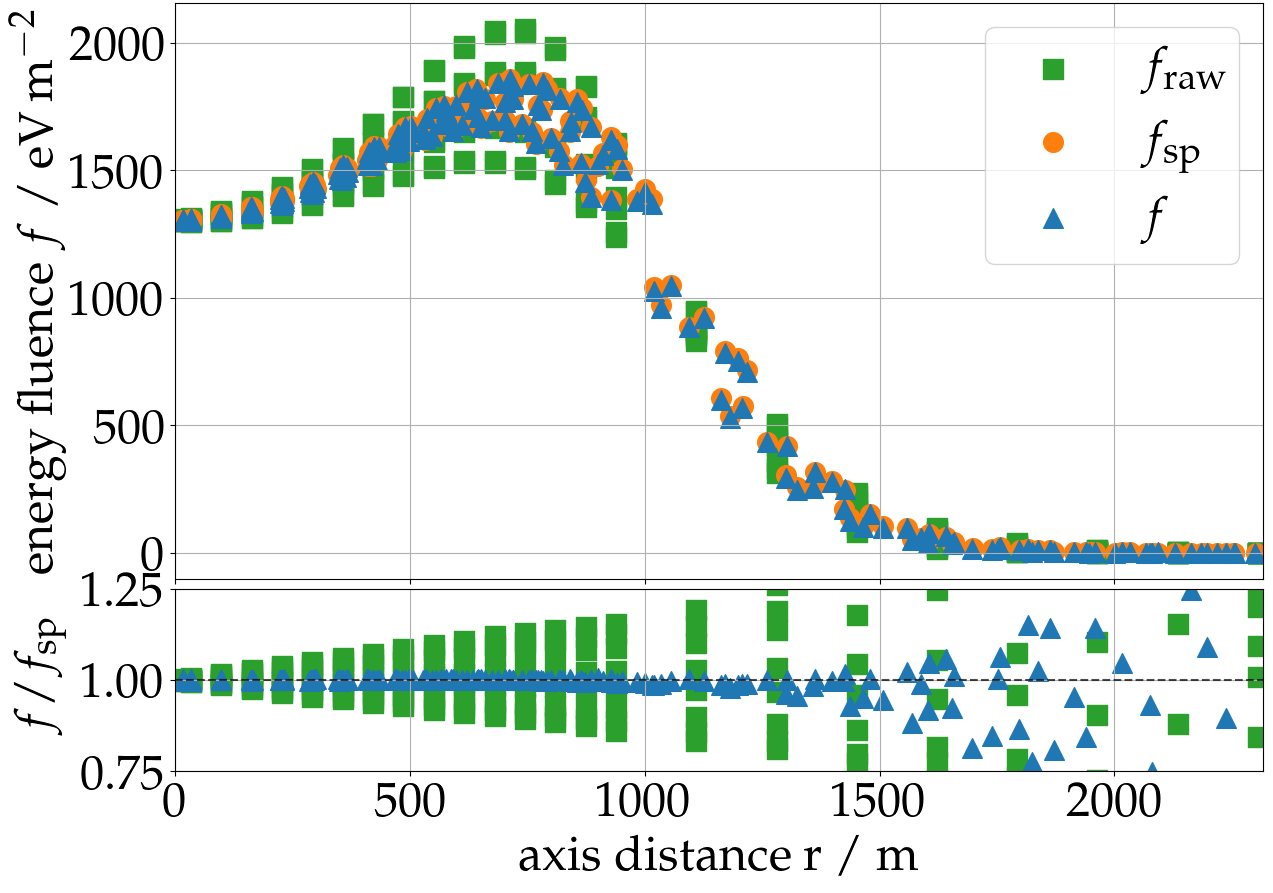}
    \caption{\textit{Right}: Early-late corrected radio-emission footprint of the same shower as in Fig.~\ref{fig:ldf:shower_ground_plane}. \textit{Right}: The uncorrected or ``raw'' (green) and corrected (blue) lateral signal distributions from observers simulated at the ground are compared to the lateral distribution from observers which are simulated in the shower plane of a shower with a zenith angle of 80$^\circ$. The bottom panel shows the relative deviation between the uncorrected and corrected ground signals to the signals simulated in the shower plane.}
    \label{fig:ldf:early-late}
\end{figure}

A description of the early-late effects has already been given in the Sec.~\ref{sec:ldf:emission} and is depicted in Fig.~\ref{fig:ldf:early-late-diagram}. To correct for these effects and eliminate the asymmetry, we ``project'' the observer positions onto the shower plane perpendicular to the shower axis intersecting with the core and thereby correct their axis distances and energy fluences. This correction assumes the radio emission to expand spherically from a point-like source at the shower maximum with the distance \dmax to the shower plane, hence the electric field amplitudes scale with the inverse of $d_\mathrm{max}$ and thus $f \sim d_\mathrm{max}^{-2}$. With this ansatz, the necessary corrections for an observer at the position $\vec{x}_i$ can be formulated with the correction factor
\begin{equation}
    \label{eq:ldf:early-late_factor}
    c_{\mathrm{el}} \equiv \frac{\dmax + \vec{x}_i \cdot \vec{e}_v}{\dmax} = 1 + \frac{z_{i}}{\dmax}.
\end{equation}
$z_i = \vec{x}_i \vec{e}_v$ is the distance between an observer and the shower plane, while the unit vector $\vec{e}_v$ points in the direction of the primary particle trajectory. With that, the corrections for the energy fluence $f$ and axis distance $r$ of individual observers are described by the following equations
\begin{equation}
    \label{eq:ldf:early-late}
    f = f_{\text{raw}} \cdot c_{\mathrm{el}}^2, \hspace{0.2cm} r = \frac{r_{\text{raw}}}{c_{\mathrm{el}}},
\end{equation}
where the subscript ``raw'' donates the uncorrected observables. Note that due to the notation of $\vec{v}$ and $\vec{B}$, observers in the positive \vvB direction are early and have a negative $z_i$ coordinate, while observers in the negative \vvB direction are late with a positive $z_i$ coordinate. Fig.~\ref{fig:ldf:early-late} (\textit{left}) shows the radio-emission footprint of the same shower as in Fig.~\ref{fig:ldf:shower_ground_plane} corrected for early-late effects using Eq.~\eqref{eq:ldf:early-late}. For this early-late corrected footprint, the symmetry w.r.t. the \vxB axis is restored as well as the overall asymmetry is decreased. This allows us in the following to describe the remaining asymmetry solely with the interference of both emission mechanisms.

To evaluate this correction, we have simulated an extra set of 17 showers which have observers on a star-shaped grid in the ground plane (equivalent to the development set) and additional observers on a star-shaped grid situated directly in the shower plane perpendicular to the shower axis. The positions of the observers in the shower plane were chosen such that they correspond to the projected, i.e., early-late correct, positions of the observers in the ground plane. In Fig. \ref{fig:ldf:early-late} (\textit{right}) the lateral distribution for such a shower with observers, simulated both in the ground and shower planes, is shown. The lateral distribution for the observers in the shower plane (orange circles) has, by definition, no early-late asymmetry imprinted and is much more narrow than the uncorrected distribution for the observers in the ground plane (green squares). The early-late corrected lateral distribution simulated at the ground (blue triangles) shows a good agreement with the distribution directly simulated in the shower plane. In the bottom panel, the ratio between the corrected ground signals and shower-plane signals shows only a slight degradation for large axis distances.
\begin{figure}[t]
    \centering
    \includegraphics[width=.7\textwidth]{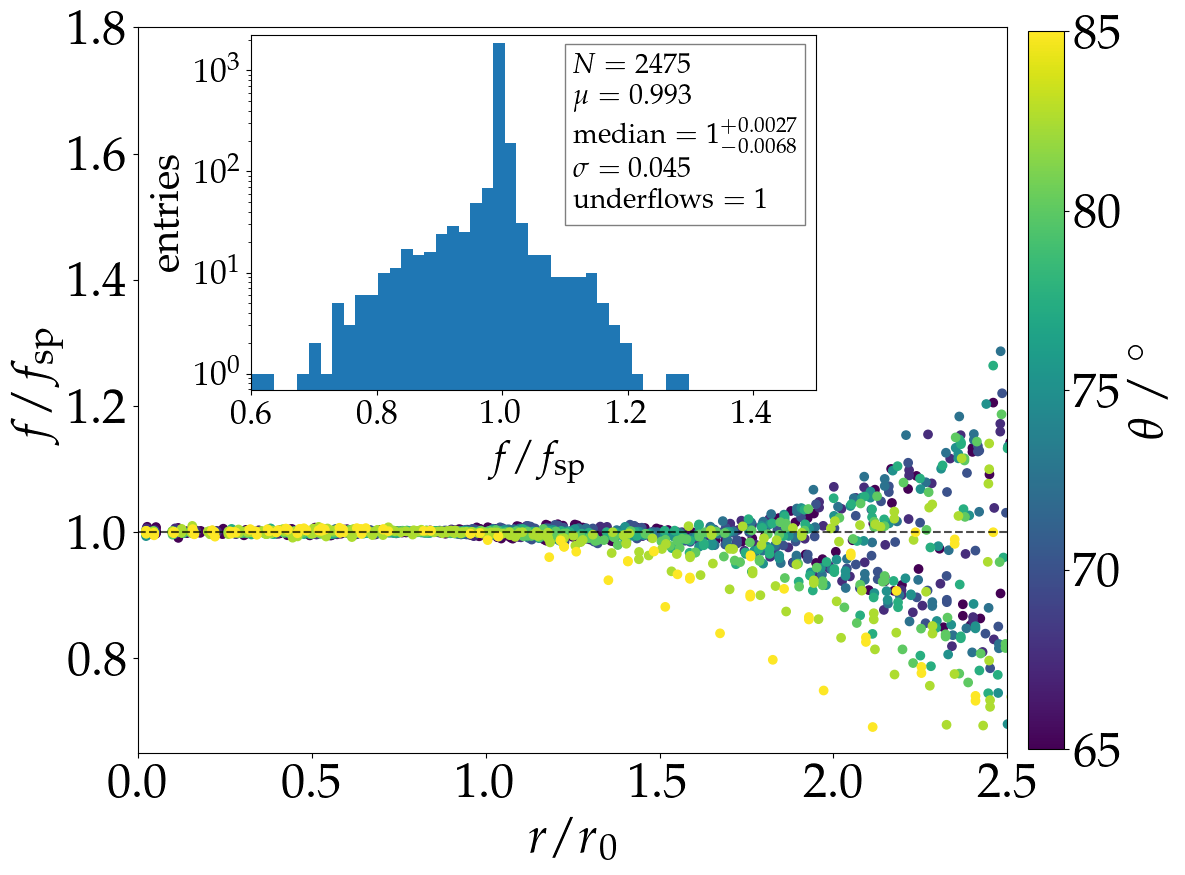}
    \caption{Relative deviation between the corrected ground signals and the signals simulated in the shower plane for all 17 showers and as a function of axis distance normalized to Cherenkov radius $r_0$, Eq.~\eqref{eq:ldf:r0}. Inset: Histogram of the relative deviation.}
    \label{fig:ldf:early-late2}
\end{figure}

A more quantitative comparison is given in Fig. \ref{fig:ldf:early-late2} which presents the ratio between corrected and directly simulated signals across 17 showers with zenith angles ranging from \SIrange{65}{85}{\degree} as a function of the lateral distance. The axis distance is normalized to the Cherenkov radius $r_0$ according to Eq.~\eqref{eq:ldf:r0}. As seen in the previous example, the accuracy decreases for larger axis distances. The inset shows a histogram of the presented data. The overall correction is better than to within 5$\,\%$.
\subsection{Lateral distribution of the geomagnetic emission}
\label{sec:ldf:geomagnetic_emission}
\begin{figure}[t]
    \centering
    \includegraphics[width=.7\textwidth]{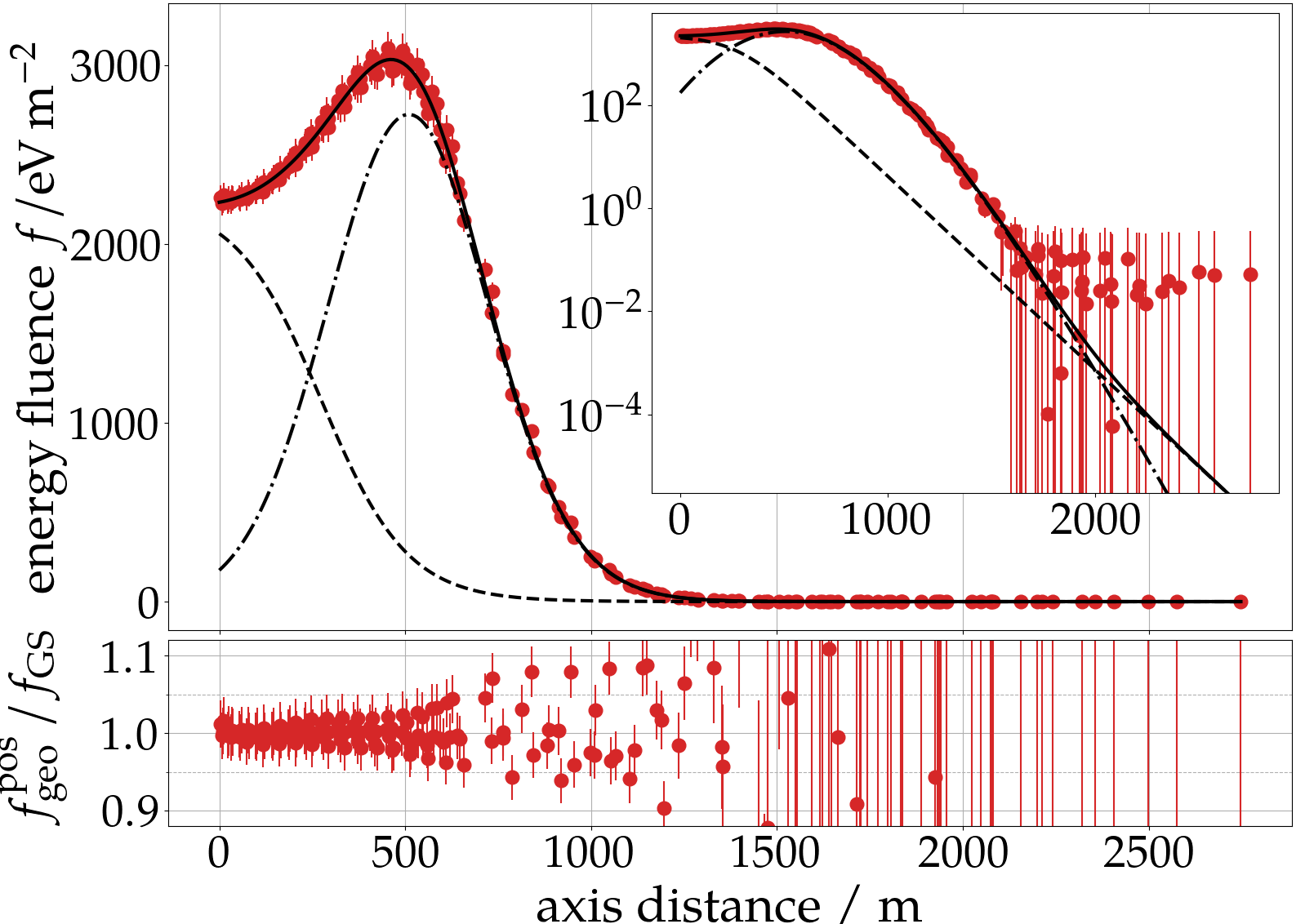}
    \caption{Lateral distribution of the geomagnetic emission from an iron-induced air shower with an energy of \eV{19.2} and a zenith angle of 75$^\circ$. The distribution is accurately described by the LDF $f_\mathrm{GS}(r)$ (solid line) which is the sum of a Gaussian (dashed-dotted line) and a sigmoid (dashed line). The inset shows the same data in log scale. The bottom panel shows the relative deviation between the markers and fitted LDF. The tail of the lateral distribution exhibits a nonphysical flattening due to thinning which is compensated for by setting appropriate uncertainties.}
    \label{fig:ldf:ldf_example}
\end{figure}
While the total radio signal exhibits asymmetries, the purely geomagnetic emission is assumed to be rotationally symmetric after any geometrical projection effects have been removed. It can thus be described by a one-dimensional LDF. In \cite{Glaser:2018byo,Bezyazeekov:2015rpa}, the LDF for the geomagnetic emission of vertical showers is modeled using a quadratic polynomial in an exponential, i.e., a Gauss curve. This allows one to describe the Cherenkov ring, i.e., the initial rise in energy fluence which is followed by an exponential decay\footnote{Such an LDF describes vertical showers only if the detector is sufficiently far from the emission region, otherwise the shower is not fully developed and the distribution of the radio emission changes. This is not a problem for inclined air showers as here detector and shower are always sufficiently far from each other.}. For more inclined showers, the Cherenkov ring increases in radius, causing a more subtle increase of the emission strength close to the shower axis. In previous iterations of our model, we used a polynomial of the 3rd order in an exponential to account for this more subtle increase. That LDF could describe the region around the Cherenkov ring accurately but decayed too rapidly at larger axis distances, undershooting the simulated signal distribution. Now, to accommodate for this phenomenon and improve the description at larger axis distances, we extend a Gaussian by the addition of a sigmoid. This yields a function \fgs with 7 parameters, an amplitude $f_0$, and 6 parameters defining the shape of the LDF $r_0^\mathrm{fit}$, $\sigma$, $p(r)$, $a_\mathrm{rel}$, $s$, and $r_{02}$:
\begin{equation}
    \fgs(r) = f_0 \left[\exp\left(-\left(\frac{r - r_0^\mathrm{fit}}{\sigma} \right) ^ {p(r)} \right)  + \frac{a_\mathrm{rel}}{1 + \exp\left(s \cdot \left[r / r_0^\mathrm{fit} - r_{02} \right]\right)} \right]. \label{eq:ldf:fgs}
\end{equation}
The Gauss-parameters $r_0^\mathrm{fit}$ and $\sigma$ can be interpreted as the position, i.e., radius, and width of the Cherenkov ring. However, it should be noted that $r_0^\mathrm{fit}$ does not coincide with the axis distance exhibiting the maximum signal strength, in fact it is slightly larger. This is plausible as the emission pattern is a superposition of the ring-like feature and a decaying exponential function. The exponent of the Gaussian, $p(r)$, is fixed to 2 for axis distances smaller than $r_0^\mathrm{fit}$ but can decrease for larger axis distances to accommodate a slower exponential decay. This allows for a better description of the tail of the LDF and was already introduced in \cite{Glaser:2018byo}. $a_\mathrm{rel}$ regulates the relative amplitude of the sigmoid term with respect to the Gauss term. The dimensionless parameters $s$ and $r_{02}$ define the shape of the sigmoid term.

Fig.~\ref{fig:ldf:ldf_example} shows the lateral profile of the geomagnetic emission of an example iron shower. The lateral profile of the geomagnetic emission is obtained by subtracting the charge-excess emission using the concept explained in Appendix \ref{app:ldf:decomposition}, i.e.,  Eq.~\eqref{eq:ldf:posfraction}, and after applying the early late correction. While fitting \fgs we allow for a shift of the core coordinates to compensate for refractive displacement. Hence, the subtraction of the charge-excess emission and early-late correction are recalculated in each iteration of the fitting procedure. For fitting we use the \textit{lmfit} python package \cite{matt_newville_2021_4516651} and a $\chi^2$ minimization. 

The lateral distribution of the geomagnetic emission is well-described by \fgs. In particular the tail (here at around \SI{1000}{m}) is more accurately described w.r.t.\ the previous iteration of our model \cite{Huege:2019cmk}. However, for even larger axis distances of around \SI{1500}{m} or more, the LDF does not follow the distribution anymore. This flattening of the simulated distribution is not expected for the coherent radio emission from extensive air showers but is rather the result of thinning, cf.\ Appendix \ref{app:ldf:thinning}. Therefore, these signals cannot be trusted. To avoid any bias in the fitting of \fgs, we use an uncertainty model for the geomagnetic energy fluence with two terms: a relative contribution of 3\%, and a constant value per shower of $10^{-4}$ the maximum geomagnetic fluence $\fgeo^\mathrm{max}$ of this shower 
\begin{equation}
    \label{eq:ldf:f_geo_uncert}
    \sigma_{\fgeo} = 0.03 \, \fgeo + 10^{-4} \, \fgeo^\mathrm{max}.
\end{equation}
The latter term ensures (relatively) large uncertainties for weak and potentially thinning-affected signals (cf.\ the large error bars in that figure). The value of $10^{-4}$ was chosen after a manual inspection of many lateral profiles, the value of 3\% was optimized to have a $\chi^2 / \mathrm{n.d.f.}$-distribution for all showers with a mean around 1.

While it is no problem to fit an LDF with 7 free parameters (+ 2 core coordinates) to a well-sampled simulated event, in experimental data the signal multiplicity is generally much lower. Furthermore, measured signals are subject to uncertainties and start values for the fit parameters are more uncertain. Hence, it is desirable to reduce the number of parameters, constrain the shape of the LDF to physically reasonable forms, and exploit correlations with shower observables. Here, we investigate the correlation of the shape parameters of Eq.~\eqref{eq:ldf:fgs} (all but $f_0$) with \dmax. It is worth stressing that this includes an implicit dependency on the zenith angle, atmospheric model, and observation height. First, we fit \fgs for all showers with a star-shaped grid. We fix the slope of the sigmoid $s = 5$ as this ensures that the sigmoid is only dominant within the Cherenkov ring, as desired, and generally simplifies the following procedure: We pick a parameter and parameterize its correlation to \dmax. Next, we fit all showers again but this time fixing the chosen parameter to its parameterization and inspecting the correlation of the next parameter with \dmax. We repeat this procedure until all parameters are described by functions of \dmax. The details of this procedure and all parameterizations are given in Appendix \ref{app:ldf:fgeo_param}.

\subsection{Parameterization of the charge-excess strength}
\label{sec:ldf:charge_excess_fraction}
\begin{figure}[t]
    \centering
    \includegraphics[width=.9\textwidth]{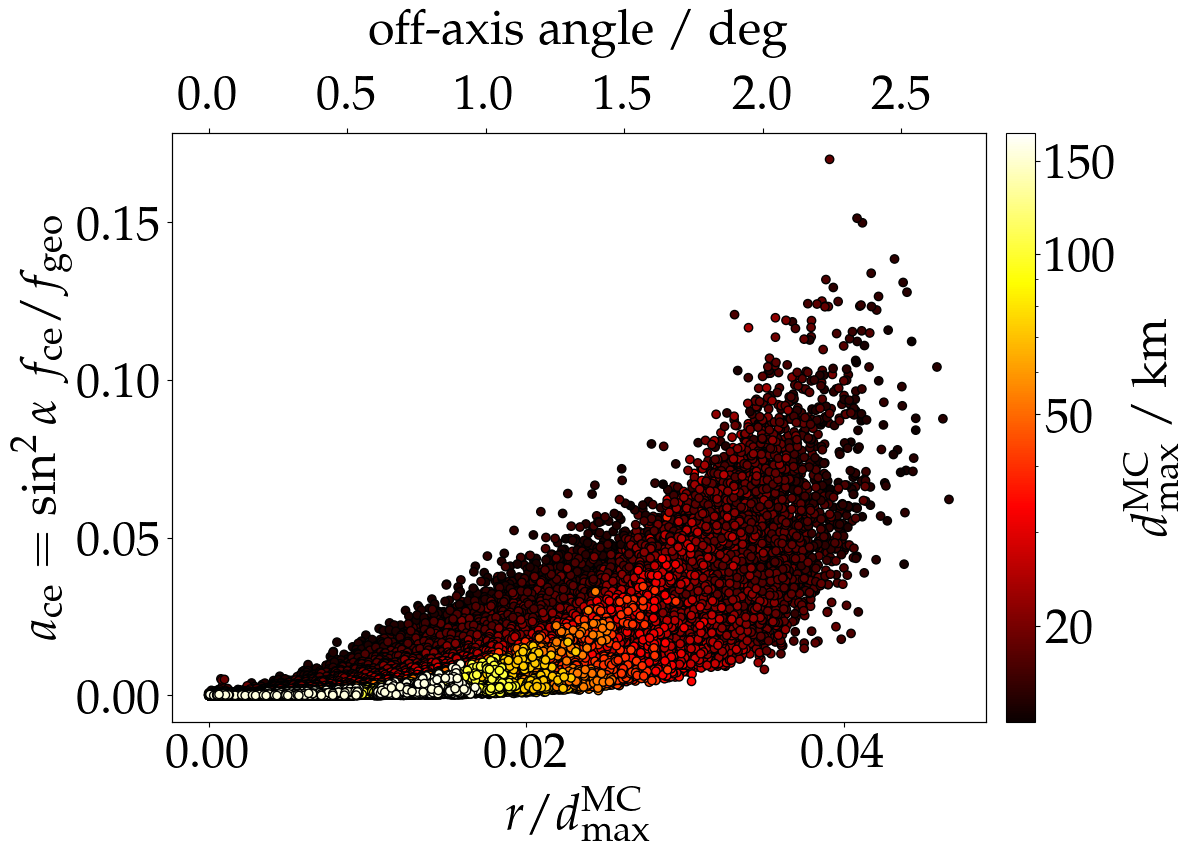}
    \caption{Lateral distribution of the charge-excess fraction \ace according to the definition \eqref{eq:ldf:charge-excess-definition} for all showers with zenith angles from 65 to 85$^\circ$. Pulses affected by thinning and on the \vB axis are excluded. The lateral distance is normalized with \dmax. For small angles this translates to the off-axis angle, the values in degree are annotated at the top. The color code shows \dmax and highlights a dependency on the shower geometry.}
    \label{fig:ldf:charge_excess_fraction}
\end{figure}
So far, we have determined the geomagnetic emission from the simulated pulses using Eq.~\eqref{eq:ldf:posfraction} based on the known polarization characteristics of both mechanisms. Thereby, the strength of the charge-excess emission which interferes with the geomagnetic emission in the \vxB polarization is estimated from the emission in the \vxvxB{-}polarization. Since the charge-excess emission and thus the emission in the \vxvxB{-}polarization is relatively weak for inclined air showers, this approach is impractical for the use with measured data as it is difficult to obtain an unbiased estimate of the true emission in the presence of ambient, thermal, Galactic, or anthropogenic noise. Hence, we follow an alternative approach where we define and parameterize a charge-excess fraction to determine the geomagnetic emission. With the following definition for the charge-excess fraction\footnote{This deviates from the definition based on amplitude ratios often used by other authors, e.g., \cite{aab2014probing}.}
\begin{equation}
    \label{eq:ldf:charge-excess-definition}
    a_\mathrm{ce} \equiv \sin^2{\alpha} \cdot f_{\mathrm{ce}} / f_{\mathrm{geo}},
\end{equation}
and Eq.~\eqref{eq:ldf:posfraction} one can derive an expression for \fgeo
\begin{equation}
    \label{eq:ldf:f_geo}
    f_\mathrm{geo} = \frac{f_{\vec{v} \times \vec{B}}}{\left(1 + \frac{\cos\phi}{|\sin{\alpha}|}  \cdot \sqrt{a_{\mathrm{ce}}} \, \right)^2},
\end{equation}
which solely depends on the (dominant) emission in the \vxB polarization. The sine of the geomagnetic angle $\alpha$, the angle between the magnetic field vector and shower axis, accounts for the scaling of the geomagnetic emission with the orientation of the shower to the magnetic field. The cosine of $\phi$, the polar angle between the observer position and the positive \vB axis, accounts for the superposition with the charge-excess emission. Using a parameterization for the charge-excess fraction and Eq.~\eqref{eq:ldf:f_geo} rather than Eq.~\eqref{eq:ldf:posfraction} has the additional advantage that the parameterization can also be used to subtract the charge-excess emission for pulses close to or on the \vxB axis. 

In the following, we use CoREAS simulations to derive a parameterization for $a_\mathrm{ce}$. First, we extract the charge-excess fraction from the simulated pulses with Eq.~\eqref{eq:ldf:posfraction}. As mentioned earlier, these equations lose validity for observers close to the \vxB axis. Hence, we only consider observers with $|\cos\phi| < 0.9$. Furthermore, we select only pulses that are not affected by thinning (cf.\ Appendix \ref{app:ldf:thinning}). In Fig.~\ref{fig:ldf:charge_excess_fraction} the lateral distribution of the charge-excess fraction for all selected pulses of all showers is shown. The lateral distance is given in terms of the off-axis or viewing angle, and \dmax is color-coded. The following behavior can be observed: First, the overall strength of the charge-excess emission decreases with increasing distance to the shower maximum, and second, it increases with the lateral distance. The former phenomenon has been studied in simulations for the total energy release between both emission mechanisms in \cite{Glaser:2016qso}, and could be shown in data as well \cite{aab2014probing}. In contrast, the charge-excess emission increases with the density and hence decreases with the zenith angle, and \dmax respectively. The scaling of the emission strength of both emission mechanisms with the density at the shower maximum \rhomax is discussed in Appendix \ref{app:ldf:emission-density-scaling}. The correlation with the lateral distance has already been reported in Refs.~\cite{Bezyazeekov:2015rpa,Schellart:2014oaa}. 
Those observations led to our ``ICRC19''-parameterization of the charge-excess fraction \cite{Huege:2019cmk}:
\begin{equation}
    a_\mathrm{ce}^\mathrm{ICRC19} = \underbrace{0.373 \cdot \frac{r}{d_{\mathrm{max}}}}_{\text{off-axis angle} \, \equiv \, p_{\mathrm{ce}, 0}} \cdot \underbrace{\exp\left(\frac{r}{762.6 \; \text{m}} \right)}_{\text{exp.\ correction} \, \equiv \, \exp\left(\frac{p_{\mathrm{ce}, 1} \, r}{1000\,\mathrm{m}}\right)} \cdot \underbrace{\left[\exp\left(\frac{\rho_{\mathrm{max}} - 0.4 \, \text{kg} \, \text{m}^{-3}}{0.149 \, \text{kg} \, \text{m}^{-3}}\right) - 0.189 \right]}_{\text{density scaling} \, \equiv \, p_{\mathrm{ce}, 2}}, \label{eq:aparam}
\end{equation}

\begin{figure}[t]
    \centering
    \includegraphics[width=.7\textwidth]{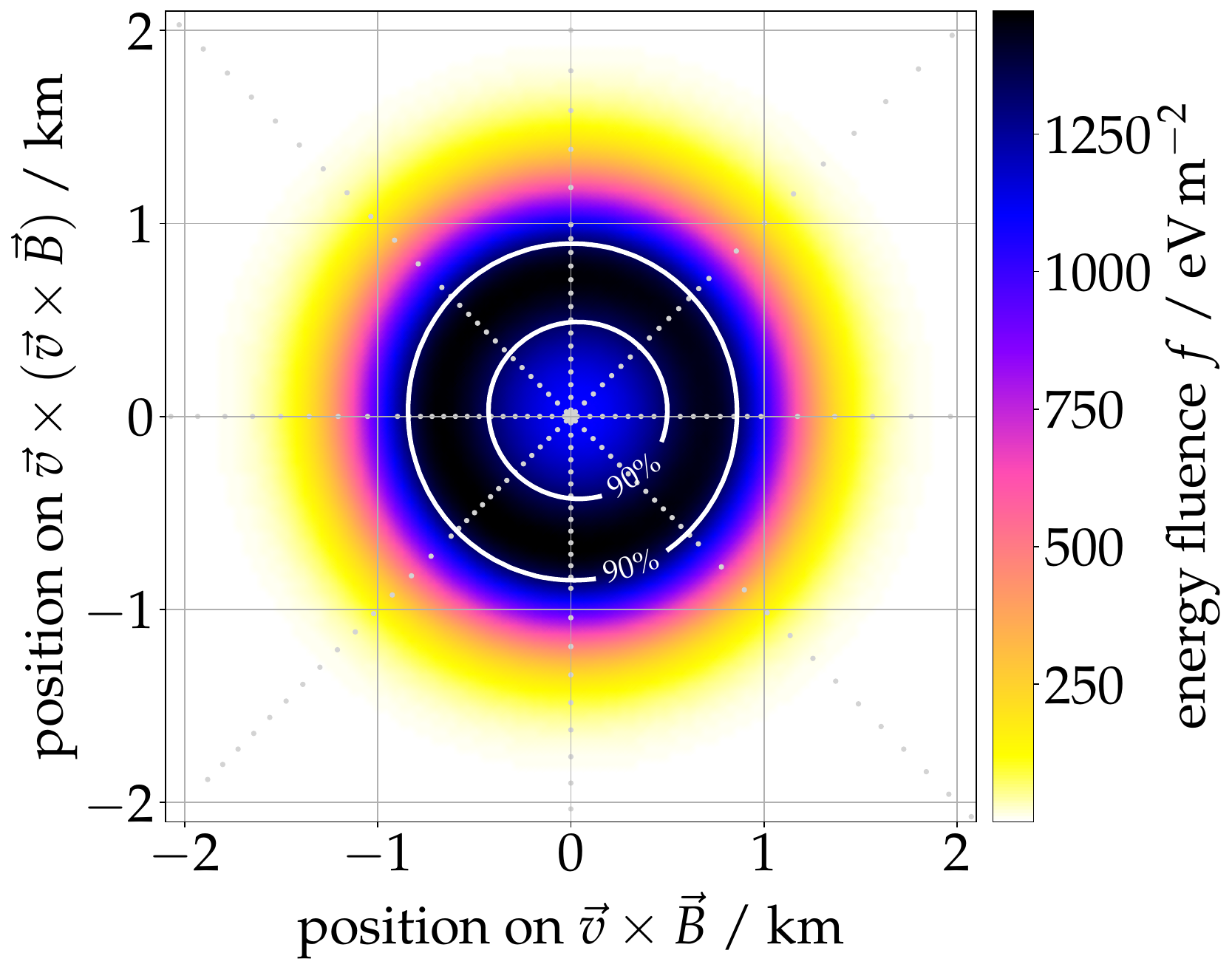}
    \caption{Footprint of the geomagnetic emission estimated using the parameterization of the charge-excess fraction \eqref{eq:ldf:charge-excess-param} and Eq.~\eqref{eq:ldf:f_geo} of the example shower also shown in Figs.~\ref{fig:ldf:shower_ground_plane} and \ref{fig:ldf:early-late} (\textit{left}). The footprint is fairly rotational symmetric around a symmetry center which is slightly displaced w.r.t. the MC shower core at the coordinate origin.}
    \label{fig:ldf:fgeo_2d}
\end{figure}

Here, we present a refined version of this parameterization. We substitute the different terms with $p_{\mathrm{ce}, i}$ $i = 0, 1, 2$, as indicated in the formula above. In an iterative procedure $p_{\mathrm{ce}, i}$ are optimized fitting Eq.~\ref{eq:ldf:f_geo} with $\ace = \ace(p_{\mathrm{ce}, i})$ to \fgs using the parameterizations established with the procedure presented in the previous section, cf.\ Appendix \ref{app:ldf:fgeo_param}. Then, the correlations of $p_{\mathrm{ce}, i}$ with \dmax, \rhomax, and $r$ are re-evaluated. The details are given in Appendix \ref{app:ldf:ce-param}. It is worth stressing that \rhomax can be determined from \dmax for a given atmospheric model and zenith angle, and thus does not introduce a new observable/fit-parameter. Finally, we can re-formulate the charge-excess fraction as a function of the $r$ and \dmax for a given zenith angle, observation height, and atmospheric model: 
\begin{equation}
    \label{eq:ldf:charge-excess-param}
    \ace = \left[0.348 - \frac{\dmax}{850.9\,\mathrm{km}} \right]  \cdot \frac{r}{\dmax} \cdot \exp\left(\frac{r}{622.3\,\mathrm{m}}\right) \cdot \left[ \left( \frac{\rhomax}{0.428 \, \mathrm{kg} \, \mathrm{m}^{-3}} \right) ^ {3.32} - 0.0057 \right].
\end{equation}
The geomagnetic emission of our example shower, estimated using this parameterization and Eq.~\eqref{eq:ldf:f_geo} with the early-late corrected energy fluence in the \vxB polarization, is shown in Fig.~\ref{fig:ldf:fgeo_2d}. Compared to the representation in Figs.~\ref{fig:ldf:shower_ground_plane} and \ref{fig:ldf:early-late} (\textit{left}), the footprint is now fairly rotational symmetric and can be described with the rotational symmetric LDF introduced in Sec.~\ref{sec:ldf:geomagnetic_emission}. Inspecting the footprint closely it becomes apparent that, although the footprint is rotationally symmetric, it is not centered around the coordinate origin which coincides with the MC shower axis. This is due to the refractive displacement of the radio-emission footprint mentioned earlier and described in \cite{Schluter:2020tdz}. 

It is worth mentioning that significant asymmetries in the lateral distribution of the charge-excess emission were reported in \cite{PhDBriechle} and attributed to shower-to-shower fluctuations. This introduces an irreducible but modest scatter of the charge-excess fraction (see evaluation in the next paragraph). On top of this, an additional dependency on the (azimuthal) arrival direction is apparent in Fig.~\ref{fig:ldf:charge_excess_fraction_param} (first 3 panels), highlighted by the color code, especially for the highest zenith angles (at which the overall relative strength of the charge-excess emission is lowest). Those characteristics are not yet understood and hence not described. They might be related to a so far unexpected dependence of the geomagnetic radiation on the orientation of the geomagnetic field vector which is shown in Sec.~\ref{sec:ldf:rec_geomagnetic_energy} and further discussed in Sec.~\ref{sec:ldf:discussion}. However, due to the low relative strength of the charge-excess emission compared to the geomagnetic emission, the remaining scatter does not significantly deteriorate the accuracy as the following evaluation shows. 

\begin{figure}[t]
    \center
    \includegraphics[width=.49\textwidth]{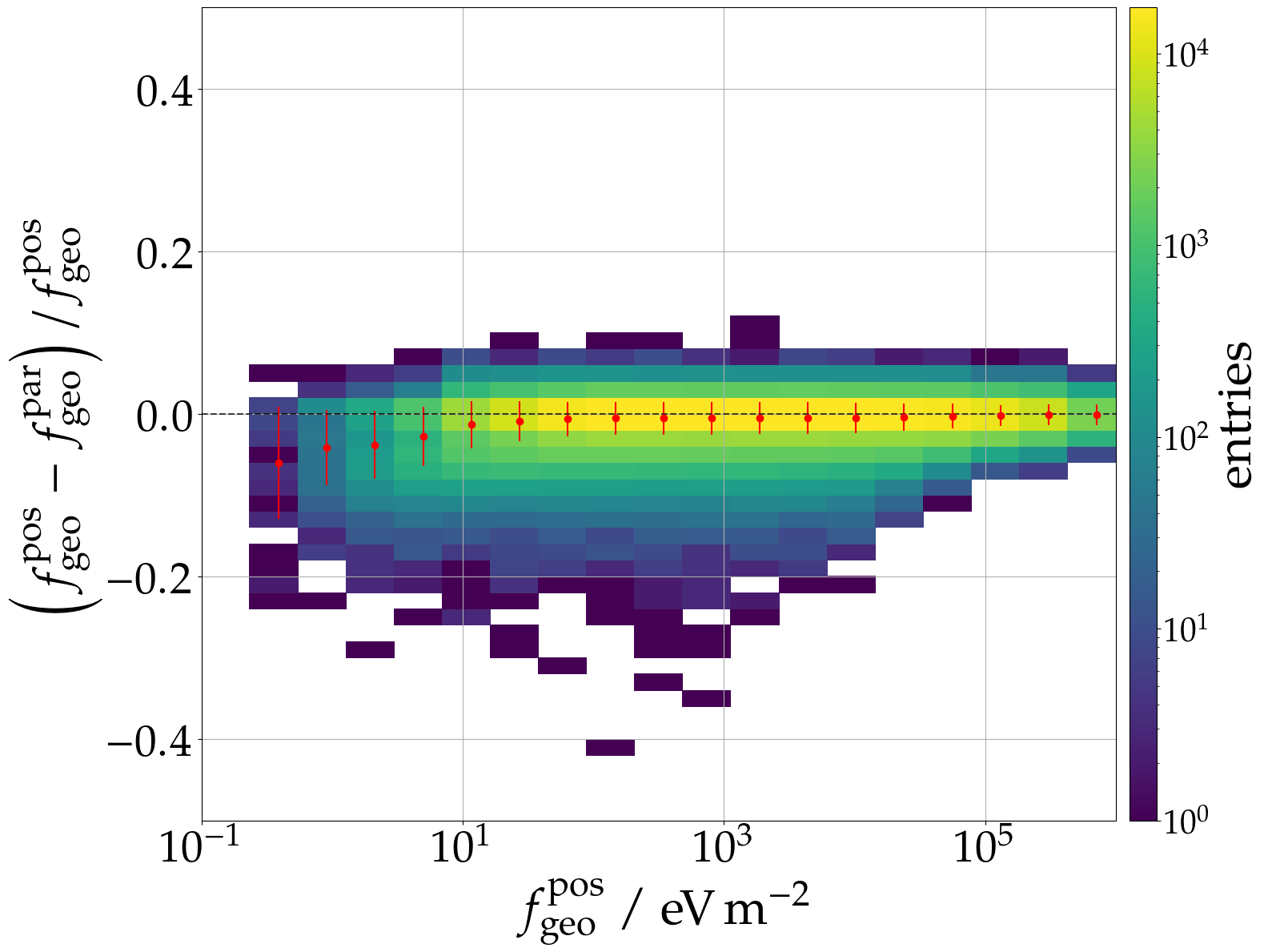}\hfill
    \includegraphics[width=.49\textwidth]{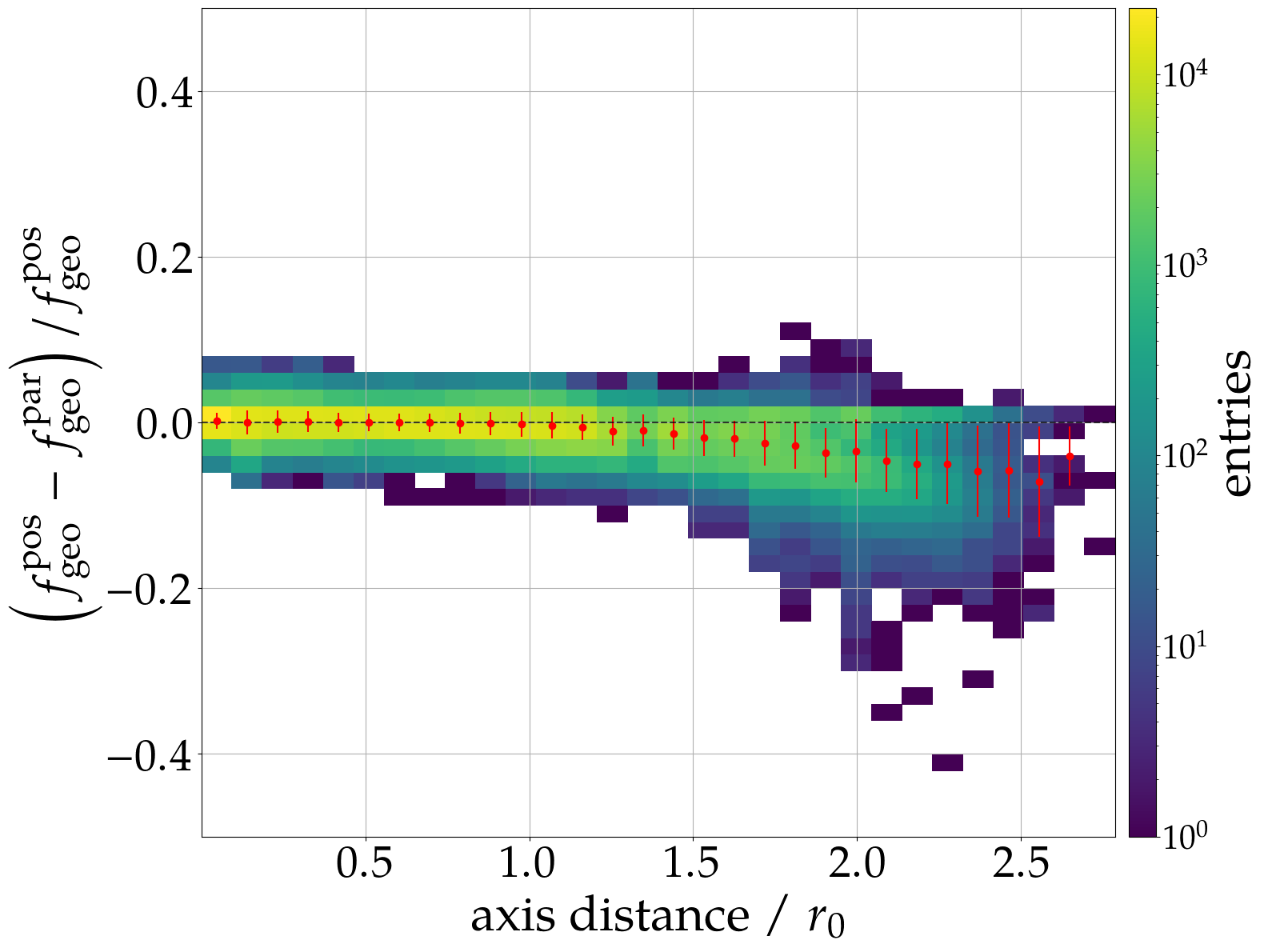}
    \caption{Comparison of the geomagnetic energy fluence $f_{\mathrm{geo}}^{\mathrm{par}}$ \eqref{eq:ldf:f_geo} determined with the parameterization of the charge-excess fraction and the geomagnetic energy fluence $f_{\mathrm{geo}}^{\mathrm{pos}}$ \eqref{eq:ldf:posfraction} calculated from the signal polarization at each simulated position. The color code shows the number of entries in each bin, the red marker show the mean and standard deviation in each vertical column. The overall agreement is better than 2\% with a mild degradation for low fluences (left panel) and distant observers (right panel).}
    \label{fig:ldf:ce_valid}
\end{figure}

To evaluate the accuracy using the parameterization~\eqref{eq:ldf:charge-excess-param}, we compare the geomagnetic energy fluence determined with Eq.~\eqref{eq:ldf:f_geo} using the parameterization~\eqref{eq:ldf:charge-excess-param}, labelled $f^\mathrm{par}_\mathrm{geo}$ for this comparison, with the fluence directly inferred from the simulated pulses using Eq.~\eqref{eq:ldf:posfraction}, labelled $f^\mathrm{pos}_\mathrm{geo}$ (par.\ = parameterization, pos.\ = position). In Figure \ref{fig:ldf:ce_valid}, the agreement between $f^\mathrm{par}_\mathrm{geo}$ and $f^\mathrm{pos}_\mathrm{geo}$ as a function of $f^\mathrm{pos}_\mathrm{geo}$ (\textit{left}) and as a function of the lateral distance from the shower axis (\textit{right}) is shown. The color code shows the number of entries in each bin on a logarithmic scale. The red markers show mean and standard deviation in each vertical column. The overall resolution is 2\% with a negligible bias. The comparison does not include observers along the \vxvxB{-}axis, where the charge excess and geomagnetic contributions are already disentangled.

\subsection{Reconstruction of the electromagnetic shower energy}
\label{sec:ldf:rec_geomagnetic_energy}
So far, we have related the shape of the signal distribution (the symmetric LDF as well as the asymmetry corrections) to \dmax. What remains is the absolute normalization $f_0$. It is easy to see that this parameter correlates with the overall emitted geomagnetic radiation energy \Egeo, the 2d spatial integral over the \fgeo footprint at the ground. We can rewrite the LDF to explicitly correlate the signal distribution to \Egeo
\begin{equation}
    \label{eq:ldf:f_geo_Egeo}
    \fgeo(r, \Egeo, \dmax) = \Egeo \frac{\fgs(r, \dmax)}{2 \pi \int_0^{5r_0} \fgs(r, \dmax) r \, \mathrm{d}r}
\end{equation}
with $f_0$ set to unity. The integral in the denominator has to be solved numerically. The maximum integration distance of $5\,r_0$ is sufficiently large to evaluate the integral without losing any significant signal. Now we can describe the entire radio-emission footprint with two fit parameters only, \Egeo and \dmax (+ two core coordinates). \Egeo is strongly correlated with the electromagnetic shower energy \Eem and hence can serve as energy estimator. It should be noted that rather than symmetrizing measured signals \fvxB by extracting the geomagnetic emission and applying the early-late correction to compare it to the geomagnetic LDF, it is more practical to apply the asymmetry correction inversely to the LDF to predict the asymmetric signal \fvxB which is directly measured and only depends on the shower arrival direction.

\section{Reconstruction of inclined air showers with a sparse antenna array}
\label{sec:ldf:rec_electromagnetic_energy}
\begin{figure}
    \centering
    \includegraphics[width=\textwidth]{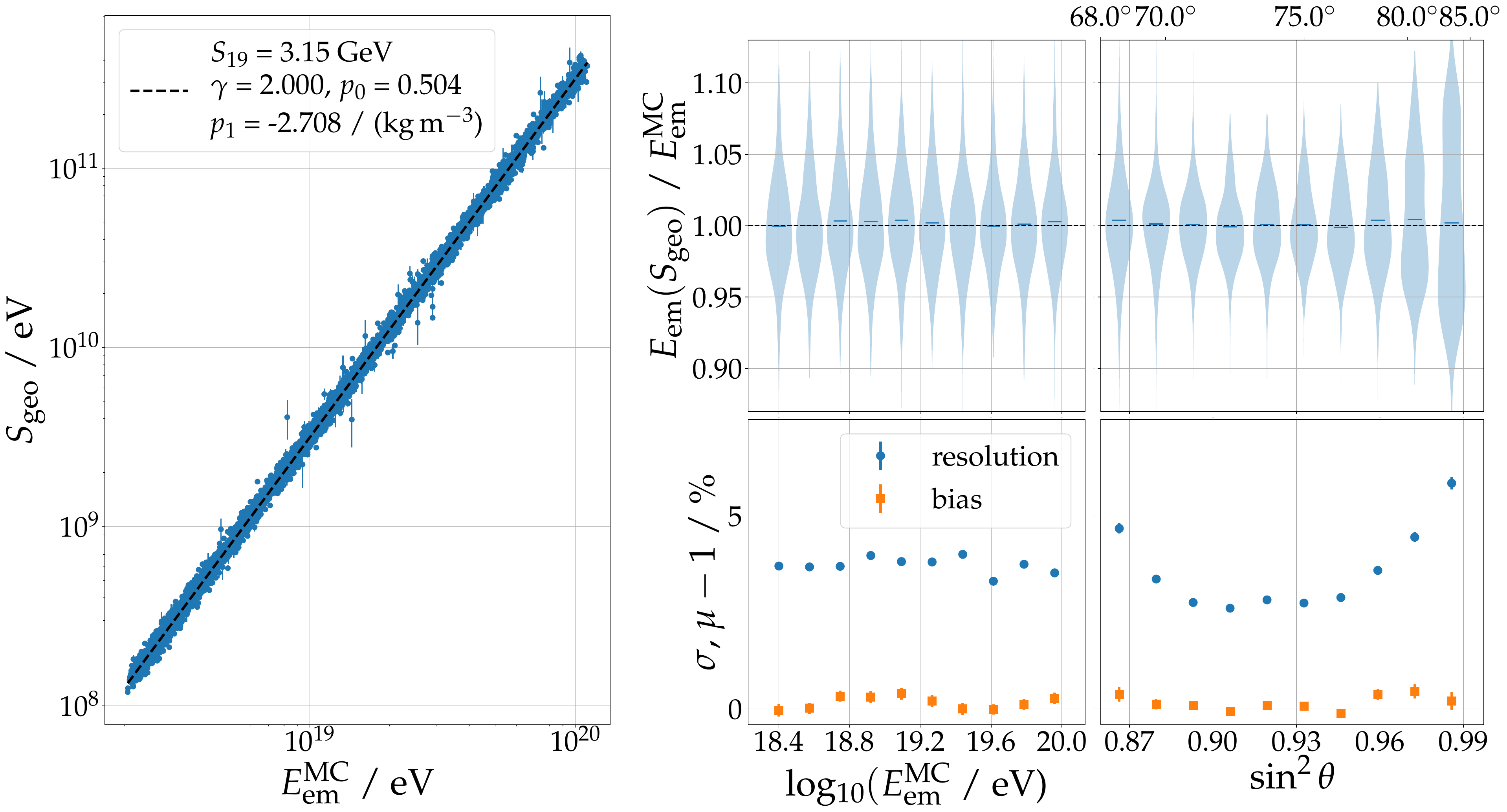}
    \caption{Reconstruction of the electromagnetic shower energy \Eem for the QGSJETII-04-generated showers. \textit{Left}: Scatter plot of the corrected geomagnetic radiation energy as function of the true electromagnetic shower energy. Legend indicates fit parameters according to Eqs.~\eqref{eq:ldf:corrS} and \eqref{eq:ldf:eem_rec}. Bias and resolution (bottom panels) of the reconstructed electromagnetic energy are shown as a function of the true energy (\textit{middle}) and zenith angle (\textit{left}). The full distributions are illustrated in the top panels.}
    \label{fig:ldf:energy}
\end{figure}

We use the second set of simulations, with the realistic \SI{1.5}{km}-spaced antenna array and continuous distributions in arrival direction and energy, to reconstruct \Egeo and \dmax with the fully parameterized signal model established in the previous section, and, in a second step, establish the correlation with the electromagnetic shower energy \Eem. For the definition of \Eem and how it is determined in simulations see Appendix \ref{app:ldf:energy}. 

For the following reconstruction, we select only QGSJETII showers with a zenith angle $\theta > 68^\circ$ and at least 5 simulated observers (no requirements on the signal strength of the simulated pulses are imposed). 6210 out of 7972 QGSJETII showers fulfill these requirements. From those 6210 showers we select 6194 showers with a good reconstruction quality. To improve the correlation with \Eem, we compensate for the second-order scaling of \Egeo with the geomagnetic angle and local air density at the shower maximum following the logic established in \cite{Glaser:2016qso}, and obtain a corrected geomagnetic radiation energy \Sgeo:
\begin{equation}
    \Sgeo = \frac{\Egeo}{\sin^2(\alpha)} \cdot
    \frac{1}{\left(1 - p_0 + p_0 \cdot \exp\left(p_1 \cdot \left[\rhomax - \langle \rho \rangle\right]\right)\right)^2},
    \label{eq:ldf:corrS}
\end{equation}
with a constant $\langle \rho \rangle = 0.3\,$g$\,$cm$^{-3}$ reflecting a typical air density at the shower maximum of an inclined air shower with $\theta \sim 75^\circ$. Finally, we can correlate \Sgeo with \Eem using a power-law:
\begin{equation}
    \label{eq:ldf:eem_rec}
    \Eem = 10\,\text{EeV} \,\, \left(\frac{\Sgeo}{S_{19}}\right)^{1 / \gamma}.
\end{equation}
The normalization with $\langle \rho \rangle$ has direct implications on the value of $S_{19}$ which can be interpreted as the geomagnetic radiation energy for a \SI{10}{EeV} cosmic ray air shower with an air density at its shower maximum of $\rhomax = 0.3\,$g$\,$cm$^{-3}$. All four parameters $S_{19}$, $\gamma$, $p_0$, and $p_1$ are determined in a combined fit of the fitted \Egeo and \rhomax (determined given the fitted \dmax) to the true \EemMC. Their values are given in Tab.~\ref{tab:ldf:s19}. The correlation between \Sgeo and \Eem (left panel) as well as the achieved reconstruction accuracy (right panels) are shown in Fig.~\ref{fig:ldf:energy}. The ratio $\Eem / \EemMC$ is shown once as a function of the true electromagnetic energy \EemMC (middle) and once as a function of the true zenith angle (right). The top panels show the full distributions in discrete bins while the bottom panels show the achieved bias ($\mu$) and resolution ($\sigma$). The reconstruction accuracy does not depend on the energy with a resolution of better than 5\% for all energies. The right panels demonstrate a minor degradation of the energy resolution for the lowest and highest zenith angles. 

\begin{figure}
    \centering
    \includegraphics[valign=t,width=0.56\textwidth]{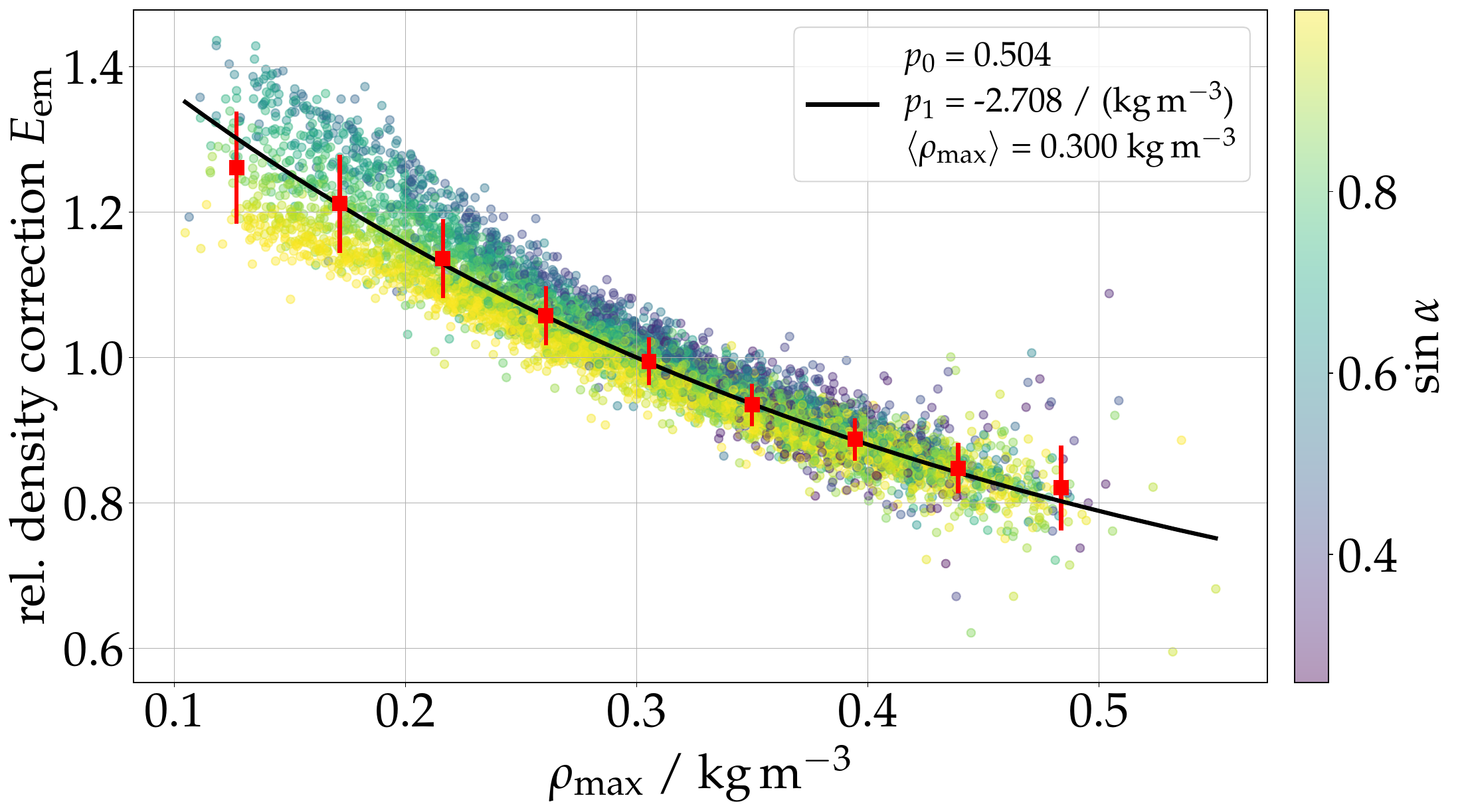}\hfill
    \includegraphics[valign=t,width=0.41\textwidth]{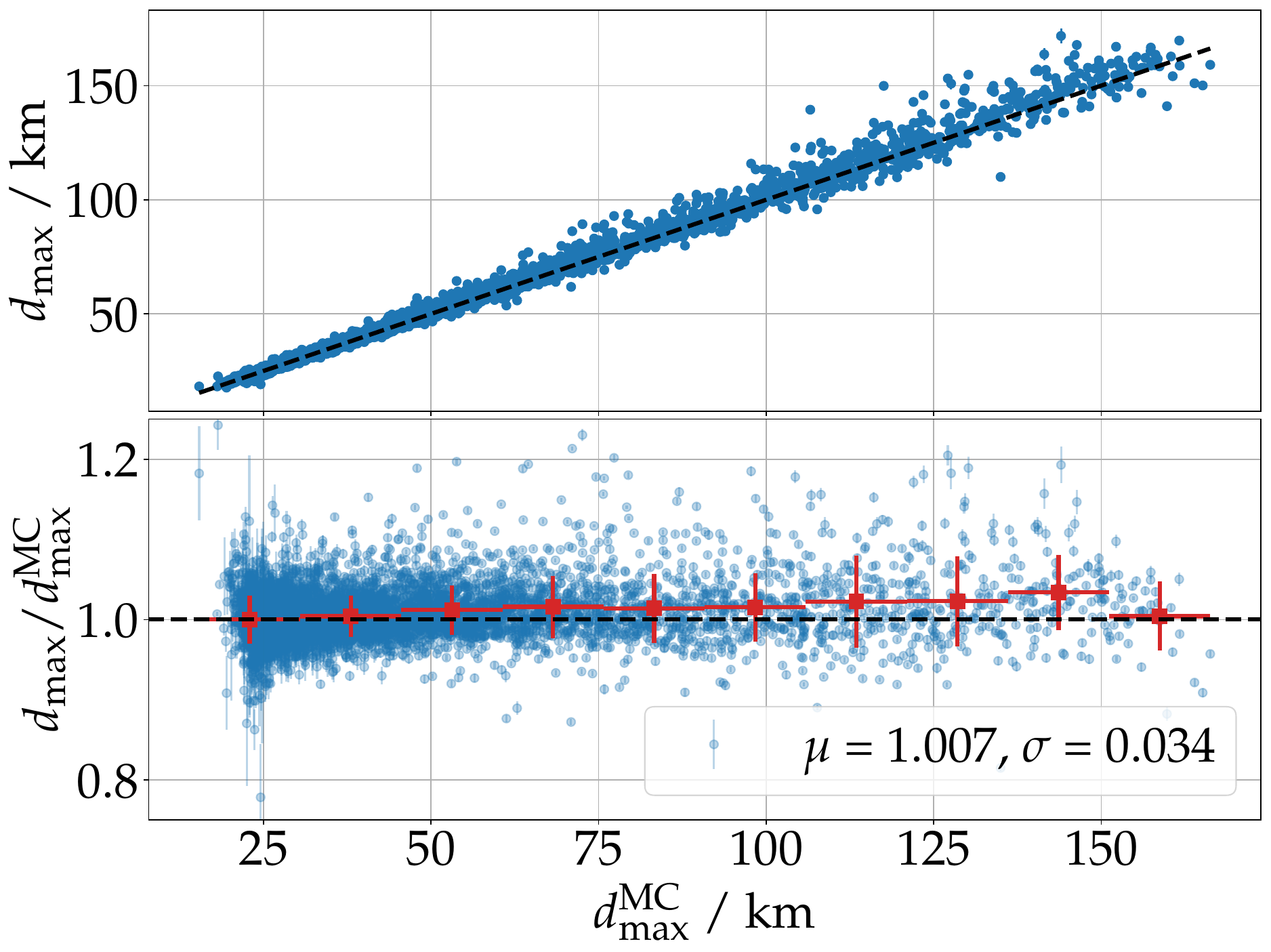}
    \caption{\textit{Left}: Fitted density correction according to Eq.~\eqref{eq:ldf:corrS} (black line) shown together with the normalized geomagnetic radiation energy from all selected showers (colored, transparent markers) together with their binned mean and standard deviation (red markers). With the normalization $y / \langle y \rangle$, $y$ according to Eq.~\eqref{eq:ldf:density_correlation}, any relative dependency to the energy and magnetic field should be removed. However, a remaining dependency on the geomagnetic angle is visible, especially for low densities (large zenith angle). \textit{Right}: Absolute (top panel) and relative (bottom panel) comparison between fitted and true distance to the shower maximum $d_\mathrm{max}$. Bias and resolution are indicated in the legend.}
    \label{fig:ldf:density}
\end{figure}

The correlation of \Eem with the air density, which can be described with the second part of Eq.~\eqref{eq:ldf:corrS}, is illustrated in Fig.~\ref{fig:ldf:density} (\textit{left}). The y-axis shows $y / \langle y \rangle$ for
\begin{equation}
    \label{eq:ldf:density_correlation}
    y = \frac{\sqrt{\Egeo / \mathrm{GeV}}}{\left(\sin\alpha^\mathrm{MC} \cdot E_\mathrm{em}^\mathrm{MC} / 10\,\mathrm{EeV}\right)},
\end{equation}
which has the dependency of the shower energy and geomagnetic angle removed. A significant correlation is visible which is well described by the fitted exponential model, i.e., the second term in Eq.~\eqref{eq:ldf:corrS}. 
The fitted model agrees well with the one found in \cite{Glaser:2016qso}, although different values for $\langle \rho \rangle$ prevent a direct comparison of the fitted parameters. The color code shows the sine of the geomagnetic angle and highlights an unexpected residual correlation which is further discussed in Sec.~\ref{sec:ldf:discussion}. This residual correlation is partially responsible for the worsening of the energy resolution at larger zenith angle.

\begin{table}[t]
    \centering
    \caption{Parameters of Eqs.~\eqref{eq:ldf:corrS} and \eqref{eq:ldf:eem_rec} determined in a combined fit.}
    \begin{tabular}{c|c|c|c}
         $S_{19}$ & $\gamma$ & $p_0$ & $p_1$ \\
         \hline
          \SI{3.1461}{GeV} & 1.9997  & 0.5045 & -2.7083 / (kg$\,$m$^{-3}$)\\
    \end{tabular}
    \label{tab:ldf:s19}
\end{table}

\begin{figure}
    \centering
    \includegraphics[width=0.7\textwidth]{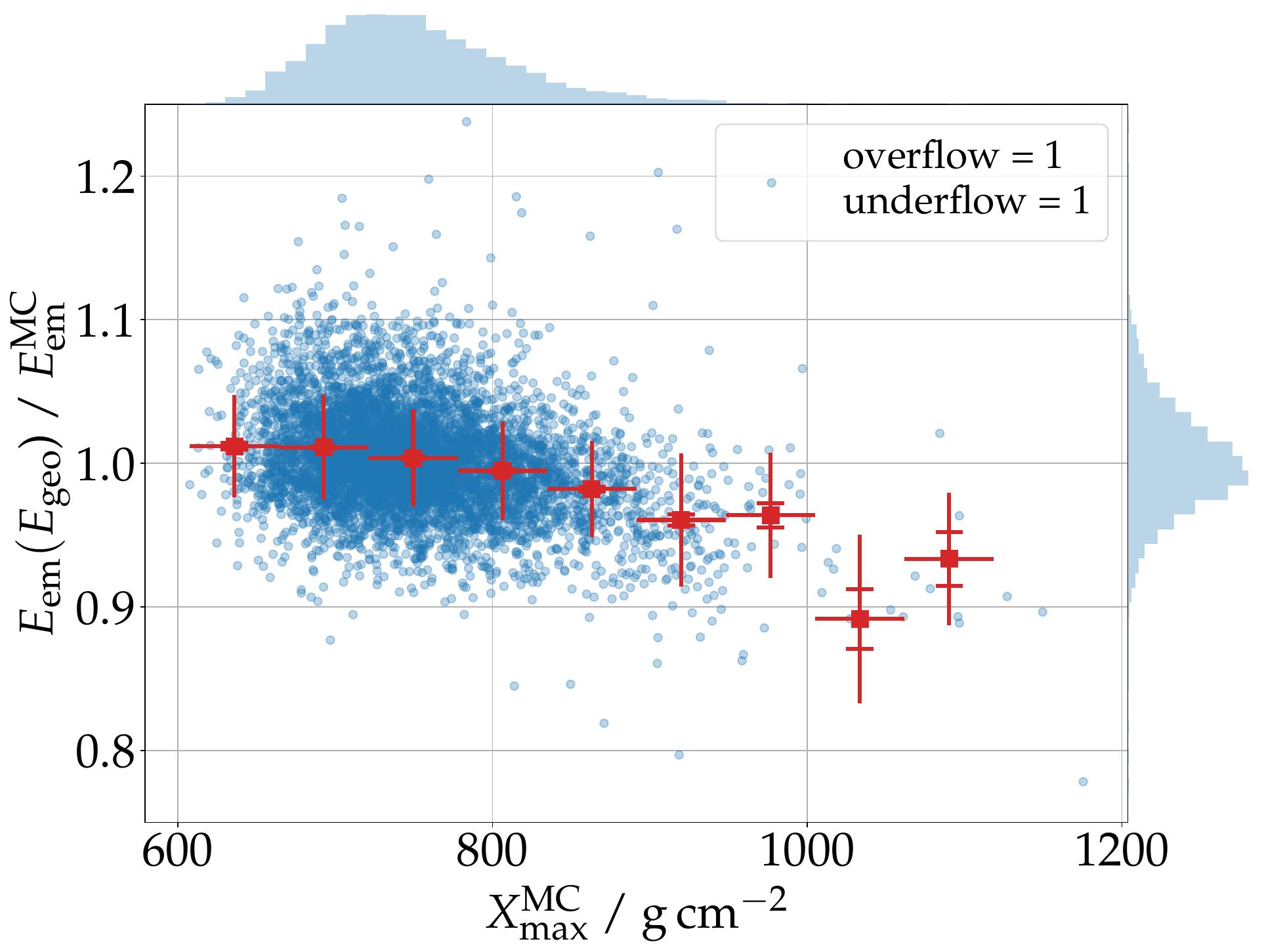}
    \caption{Scatter plot of $E_\mathrm{em}(\Egeo) / E_\mathrm{em}^\mathrm{MC}$ as a function of $\Xmax^\mathrm{MC}$. Red markers show mean and standard deviation (error-caps signify uncertainty of the mean). A \Xmax dependent bias is apparent. The distributions of $\Xmax^\mathrm{MC}$ and $E_\mathrm{em} / E_\mathrm{em}^\mathrm{MC}$ are illustrated by the histograms on the top and right sides of the panel, respectively.}
    \label{fig:ldf:xmax_bias}
\end{figure}

Fig.~\ref{fig:ldf:xmax_bias} shows the ratio $\Eem / \EemMC$ as a function of the true $\Xmax^\mathrm{MC}$ for each shower (blue dots). The binned mean and standard deviation (error bars) are highlighted by the red markers, the uncertainties on the means are indicated by the error caps. A bias with $\Xmax^\mathrm{MC}$ is visible: for larger $\Xmax^\mathrm{MC}$, the reconstructed electromagnetic energy is underestimated. The overall distributions of $\Xmax^\mathrm{MC}$ and $\Eem / \EemMC$ are shown as histograms at the top and right axes, respectively. A potential \Xmax-dependent bias in the energy reconstruction is delicate as it could yield a primary-particle dependent bias. However, more than 97\% of events are contained within $\Xmax^\mathrm{MC} < 900\,\gcm$ for which the bias is below 5\%\footnote{The \Xmax-distribution depends of course on the energy spectrum of the simulation set. The simulated spectrum is much harder than what is seen in nature, hence larger energies, i.e., deeper shower, are over-represented in our simulations.}. Furthermore, we did not observe any  significant bias in the electromagnetic energy reconstruction between the different primaries. Nonetheless, in a future iteration of this reconstruction this could be improved as discussed in Sec.~\ref{sec:ldf:discussion}.

\subsection{Reconstruction of the distance to the shower maximum}
In Fig.~\ref{fig:ldf:density} (\textit{right}), the reconstructed \dmax is compared to its true value. The comparison shows an overall good accuracy with no significant bias and a resolution of $\sim \,3\%$ which does not significantly depend on the zenith angle. It should be mentioned that the superb reconstruction accuracy of \dmax achieved here is mainly driven by the fact that we are using the true arrival direction in the reconstruction. In a realistic, experimental setup, where the arrival direction for inclined air showers is only known with a typical accuracy of $\lesssim \SI{0.5}{\degree}$ \cite{Aab:2014gua}\footnote{The detection of the air-shower front from radio emission does not suffer from Poisson fluctuation (as it is the situation with the detection of particles), hence a more accurate reconstruction of the shower arrival directions is likely possible.}, the reconstruction accuracy of \dmax will decrease, while we expect that the accuracy on \Eem is only marginally effected by this. The potential to use \dmax to reconstruct \Xmax is discussed in Sec.~\ref{sec:ldf:discussion}.

\subsection{Reconstruction of air showers generated with a different high-energy hadronic interaction model}
We repeated the same evaluation of the air-shower reconstruction with a sparse, realistic antenna array with the Sibyll2.3d-generated air showers. To reconstruct the electromagnetic energy \Eem, the parameters $\gamma$, $p_0$, and $p_1$ are fixed to the values obtained with the QGSJETII showers to allow a direct comparison of the $S_{19}$ parameter. $S_{19}$ decreased by less than 2\% from \SI{3.15}{GeV} to \SI{3.10}{GeV} for the Sibyll showers as compared with the QGSJETII showers. It is worth stressing that this change is not due to differences in the prediction of the muonic shower component between both hadronic interaction models, because we reconstruct the electromagnetic energy of the shower. The achieved resolution is very comparable between showers from both interaction models with the exception that the resolution for air showers with zenith angle $\theta < \SI{70}{\degree}$ for Sibyll showed a small degradation in resolution ($\sigma_{\Eem}^\mathrm{Sibyll}(\theta < \SI{70}{\degree}) \lesssim 7\%$ as compared to $\sigma_{\Eem}^\mathrm{QGSJET}(\theta < \SI{70}{\degree}) \lesssim 5\%$). See Fig.~\ref{fig:ldf:sibyll} in Appendix \ref{app:ldf:energy-sibyll}. The other results, e.g., the reconstruction of \dmax and the \Eem reconstruction bias with \Xmax remain practically unchanged. 

\section{Discussion}
\label{sec:ldf:discussion}
In this section, we elaborate on several features of the here presented signal model and energy reconstruction. In particular, we address the question how the signal model and established parameterizations adapt to different ambient conditions and frequency bands.


The model presented here for the radio emission from inclined extensive air showers in the frequency band of \SIrange{30}{80}{MHz} is tailored to the ambient conditions of the Pierre Auger Observatory. While the general concept and considerations should transform well to other experiments, i.e., other ambient conditions and frequency bands, like GRAND, the explicit parameterizations require revisions. For example, it is known that the Cherenkov ring is more prominent at higher frequencies \cite{Schluter:2020tdz}, hence a re-parameterization of the shape of the 1-d LDF seems necessary. Relying on atmospheric models for the parameterization of the charge-excess emission and parts of the LDF parameters reduces the dependency on a particular set of ambient conditions. However, the explicit use of the distance to shower axis $r$ in the parameterization of the charge-excess fraction \eqref{eq:ldf:charge-excess-param} and LDF model \eqref{eq:ldf:fgs} carries a dependency on the observation altitude. 

We normalized our parameterizations to showers arriving perpendicular to the Earth's magnetic field, hence changing orientations of the magnetic field should not affect the model. In \cite{Glaser:2016qso}, the scaling of the (geomagnetic) emission with the strength of the Earth's geomagnetic field was investigated and found to be $\Egeo \sim B^{1.8}$. This scaling should apply to our model as well.

However, it should be noted, that in Fig.~\ref{fig:ldf:density} a residual correlation with $\sin\alpha$ is apparent which is not yet understood. This correlation becomes stronger with decreasing air density. In \cite[Fig.\ 2]{Chiche:2021iin} the density scaling of the geomagnetic emission was also investigated but with a stronger geomagnetic field and for the radio emission at higher frequencies. A strong suppression of the geomagnetic emission at lower densities was found, while the behavior at larger air densities is in agreement with the results shown here and in \cite{Glaser:2016qso}. While we do not see such a suppression here, a common causation related to the magnetic field strength, i.e., the magnitude of the Lorentz force, seems reasonable. In \cite{James:2022mea}, a transition for the geomagnetic radio emission from the regime of time-varying transverse currents to a regime of synchrotron radiation is predicted for air showers developing in low air density in the presence of strong magnetic fields. We do not observe a clear transition in the phase space covered by our simulations (as it is also not expected), however the residual correlation with $\sin\alpha$ as well as the suppression of the emission for very low air densities reported in \cite{Chiche:2021iin} might be explained by this transition. If this explanation proves to be accurate, which needs further investigations, the model presented here requires an adaptation for sites with stronger magnetic fields than present at the site of the Pierre Auger Observatory.

The reconstruction of the electromagnetic energy presented here corresponds to an idealized case and hence the achieved resolution can be considered the intrinsic resolution of the method for air showers reconstructed with a sparse antenna array. In measured data, neither the true arrival direction nor the exact signals arriving at the observers are known perfectly. Ambient and internal noise, an inaccurate detector description (especially of the directional response pattern of the antennas), and other effects will affect the signals reconstructed for each antenna. A detailed study of all those effects is beyond the scope of this paper. A more realistic study of the achievable reconstruction accuracy and reconstruction efficiency with this model has been conducted in \cite{PierreAuger:2021ece}.

In addition to its application in an event reconstruction, this signal model can also be used to predict the radio emission in the \SIrange{30}{80}{MHz} band from inclined air showers of a given set of energies and arrival directions. This allows studying different aspects of the detection of inclined air showers, for example the effect of the observer spacing on the detection efficiency, when time- and CPU-consuming Monte-Carlo simulations are not available.

Besides the electromagnetic shower energy, the distance of the shower maximum is an observable of great interest as it can be used to determine the slant depth of the shower maximum \Xmax. \Xmax is of special interest as it is commonly used to infer the mass composition of cosmic rays. The distance to the shower maximum \dmax is reconstructed with a superb accuracy of $\sigma_{\dmax} / \dmax = 3\%$ as shown in Fig.~\ref{fig:ldf:density} (\textit{right}). While we intentionally do not probe a realistic scenario, it is worth mentioning (again) that the true arrival direction (zenith angle) is used, with which \dmax is mostly correlated. Furthermore, small relative changes in \dmax correspond to large (absolute) changes in \Xmax. Even with a relative \dmax resolution of $3\%$, the absolute resolution for the depth of the shower maximum is $\sigma_{\Xmax} \geq 50\gcm$ at $\dmax \geq 75\,$km. This leads to the conclusion that the sensitivity of the shape of the lateral signal distribution at ground to \Xmax is rather limited for inclined air showers due to the large distance between shower maximum and detector and the (relatively) small variations in \dmax induced by variations of \Xmax. While this is unfortunate for obvious reasons, when one wants to estimate the cosmic ray energy it is advantageous as it minimizes the dependency of the LDF to the mass of the cosmic ray primary.  

\section{Conclusions}
\label{sec:ldf:conclusiuon}
Measuring inclined air showers with radio antennas is of particular interest for two reasons. First, their large footprints allow us to instrument huge areas with sparse antenna arrays, which are necessary to observe the spectrum of cosmic rays at the highest energies. Second, inclined air showers observed in coincidence with radio and particle detectors offer the unique potential to measure the muonic shower component (with the particle detector) and electromagnetic shower component (with the radio detector) independently of each other. The combination of this complementary information yields a strong sensitivity towards the mass of the cosmic ray. For a precise study of the mass composition of UHECR, the energy resolution provided by the radio detector is of critical importance. We present a signal model for the radio emission in the \SIrange{30}{80}{MHz} band from inclined air showers. The model enables accurate reconstruction of the electromagnetic shower energy with a sparse radio-antenna array, an intrinsic resolution of better than 5\% and no bias ($< 1\%$) on the primary particle mass. The model relies on an explicit modeling of the dominant, rotationally symmetric geomagnetic emission as well as effects which disturb this symmetric emission and lead to the highly asymmetric pattern we expect from inclined air showers. Those asymmetries are associated with the interference of the charge-excess emission with the geomagnetic emission as well as the imprint of geometrical early-late effects. We exploit correlations between the model parameters and shower observables to minimize the number of free parameters. The final model only relies on two free parameters, the distance between detector and the shower maximum \dmax and the geomagnetic radiation energy \Egeo, plus two coordinates for the location of the impact point of the air shower. This allows a reliable fit of the signal distribution and thus efficient event reconstruction.

The presented concept for the signal model is applicable for a variety of radio experiments trying to reconstruct inclined air showers. The described procedure can be used to tune the model parameterizations to match with different ambient conditions as well as different frequency bands relevant for a specific experiment.

\acknowledgments
We would like to thank Alan Coleman for his suggestion to use a sigmoid in addition to a Gauss to describe the lateral profile of the geomagnetic emission as well as for his comments to our manuscript. Also, we would like to thank the anonymous reviewer whose detailed comments helped to greatly improved the quality of this paper. Furthermore, we are thankful to our colleagues involved in radio detection within the Pierre Auger Observatory for very fruitful discussions. We thank our colleague M. Gottowik for his contribution to the simulation library with the PLEIADES cluster at the University of Wuppertal. Felix Schlüter is supported by the Helmholtz International Research School for Astroparticle Physics and Enabling Technologies (HIRSAP) (grant number HIRS-0009). Simulations for this work were performed on the supercomputer ForHLR and its successor HoreKa at KIT and the PLEIADES cluster at the University of Wuppertal. ForHLR and HoreKa are funded by the Ministry of Science, Research and the Arts Baden-Württemberg and the Federal Ministry of Education and Research. PLEIADES is supported by the Deutsche Forschungsgemeinschaft (DFG).

\appendix


\section{Treatment of simulated signals}

\subsection{High-frequency emission artifacts from particle thinning}
\label{app:ldf:thinning}
To compute the radio emission from (inclined) air showers with reasonable computational effort, a technique called \textit{thinning} is used \cite{Kobal:2001jx}. This implies that particles produced in a single interaction and below a certain energy threshold are removed from the simulation except for one randomly selected particle. This particle is assigned a weight to describe the ensemble of particles it ``replaces'' such that energy conservation is preserved. The probability for a particle to be selected is proportional to its energy. This dramatically reduces the number of particles to be simulated while correctly reproducing showers on average. Random particle fluctuations and thus shower-to-shower fluctuations are affected. However, if the energy threshold $E_\mathrm{th} = \epsilon_\mathrm{thin} E_0$ and the maximum weight a particle can be assigned $w_\mathrm{max} = \epsilon_\mathrm{thin} E_0 / \text{GeV}$ which both depend on the parameter $\epsilon_\mathrm{thin}$, are chosen wisely \cite{Kobal:2001jx}, the effect is tolerable. For the simulation of the radio emission, \textit{thinning} introduces another problem. A particle with a large weight, which represents many particles, emits a radio wave with an amplitude scaling with its weight (with no phase difference) while the actual ensemble of particles emits radio waves with phase differences. In other words, particles which are described by one particle with a corresponding weight emit perfectly coherent emission. This effectively introduces artificial additional power. For small lateral observer distances, this power is well below the actual coherent radio emission. However, for increasing lateral distances or when considering higher frequencies, i.e., with decreasing coherence, this artificial signal starts to significantly impact the simulated power and subsequently, the affected pulses need to be rejected.

\begin{figure}[t]
    \includegraphics[width=\textwidth]{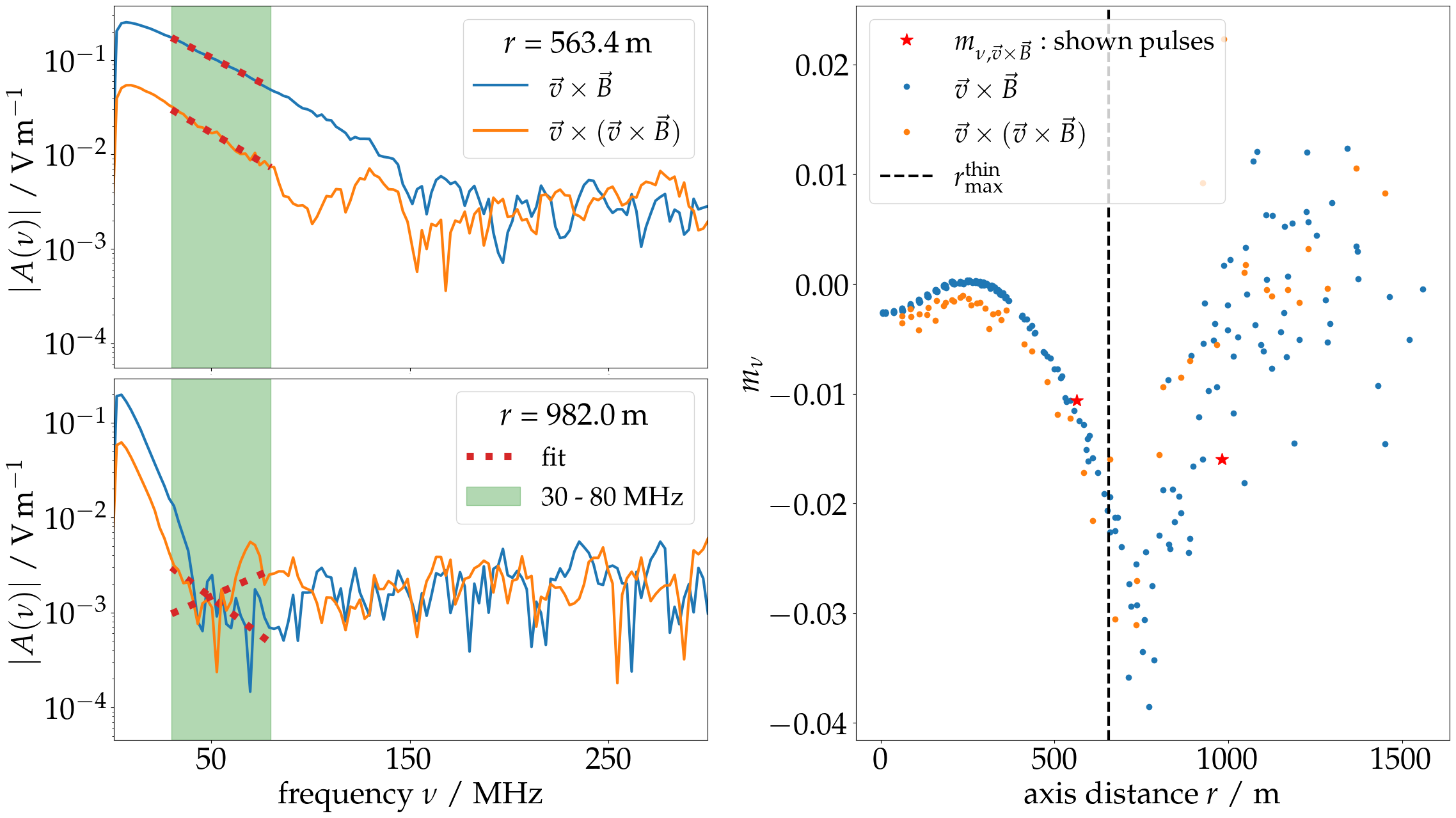}
    \caption{\textit{Left:} Absolute amplitude spectrum as function of the frequency $|A(\nu)|$ of two simulated pulses of observers situated along the $\vec{v} \times (\vec{v} \times \vec{B})$ axis at different distances to the MC shower axis (cf.\ legends). The spectra of the $\vec{v} \times \vec{B}$ (blue) and $\vec{v} \times (\vec{v} \times \vec{B})$ (orange) polarizations (i.e., geomagnetic and charge-excess emission for observers along the $\vec{v} \times (\vec{v} \times \vec{B})$ axis) are shown individually. The greenish band highlights our frequency band of interest, the dashed red lines show fits according to Eq.~\eqref{eq:ldf:thinning}. \textit{Right:} The fitted slope $m_\nu$ as function of the lateral distance for the two different polarizations. For the $\vec{v} \times (\vec{v} \times \vec{B})$ polarization, only observers on the $\vec{v} \times (\vec{v} \times \vec{B})$ axis and with an axis distance of at least \SI{50}{m} are shown, because pulses close to the shower axis or along the $\vec{v} \times \vec{B}$ axis contain almost no signal in the $\vec{v} \times (\vec{v} \times \vec{B})$-polarization. The slopes in the $\vec{v} \times \vec{B}$ polarization from the pulses presented in the left panels are highlighted with stars. The dashed, black line indicates the distance cut used to identify pulses affected by thinning, see details in the text.}
    \label{fig:ldf:thinning}
\end{figure}

In the left panel of Fig.~\ref{fig:ldf:thinning}, the spectra of two pulses are presented. The observer of one pulse is closer to the shower axis (\textit{top}) and the other one further away (\textit{bottom}). For both pulses, the spectra of the \vB{-} and \vvB{-}polarizations are shown, representing the signals of the geomagnetic and charge-excess emission contributions, respectively, as both observers are situated along the \vvB axis. The band of interest from \SIrange{30}{80}{MHz} is highlighted. Both spectra show the same feature: A smooth exponential decay of the amplitude followed by a noisy plateau. While the first is expected for coherent emission, the latter is not and thus interpreted to be caused by thinning. While the pulse of the closer observer is not (or in the case of the \vvB{-}polarization only slightly) affected by the noise floor in the band of interest, the pulses of the observer further away from the shower axis show a significant disruption in both polarizations and thus have to be rejected from further analysis. To quantitatively examine whether a pulse is contaminated or not, we fit a first-order polynomial to the logarithmic spectrum in the frequency $\nu$ range between \SIrange{30}{80}{MHz}, i.e.,
\begin{equation}
    \label{eq:ldf:thinning}
    |A(\nu)| = 10^{m_\nu \cdot \nu + b}
\end{equation}
with a slope parameter $m_\nu$ and a constant $b$. The slope parameter $m_\nu$ as a function of the lateral distance for an example shower is shown in Fig.~\ref{fig:ldf:thinning} (\textit{right}). While the spectrum is almost flat ($m_\nu = 0 $) on and around the Cherenkov ring, it is falling more steeply with increasing lateral distance as expected. Around 750$\,$m a kink is visible. The lateral distance of the observer whose pulse in the \vxB polarization has the steepest slope after which the disruption in the considered band becomes considerable, defined as $r_\text{min}$, is identified. To be conservative, we select pulses only from observers with a lateral distance smaller than $r_\text{min}^\text{thin} = 0.85\, r_\text{min}$ as clean, pulses of observers with larger lateral distances are considered affected by thinning artifacts. For the example event, the dashed line indicates this criterion. The considered maximum lateral distance per shower scales in first order with the zenith angle and just slightly with the energy. This is expected since energy-dependent weight limitation was used \cite{Kobal:2001jx} to simulate the air showers. For highly inclined showers, observers with lateral distances of over 2 km are still considered. With this selection, the number of considered observers decreases from 240 to around 160 per simulated shower. This selection is solely used for the parameterization of the charge-excess fraction in Sec.~\ref{sec:ldf:charge_excess_fraction} given the fact that it is otherwise difficult to independently identify affected observers and mitigate their effect on the parameterization. For fitting the lateral distribution of the geomagnetic emission (cf.\ Sec.~\ref{sec:ldf:geomagnetic_emission}), we consider all observers - even the ones with pulses affected by thinning artifacts - but assign an appropriate uncertainty to all signals, effectively reducing the impact of weak signals, to avoid any bias from affected observers. 

\subsection{Decomposition of the radio signal}
\label{app:ldf:decomposition}
For an accurate description of the radio-emission footprints it is useful to decompose the emission into the geomagnetic and charge-excess contributions which is possible due to their polarizations characteristics. The geomagnetic emission is polarized in the negative \vxB-direction while the charge-excess emission is polarized radially inwards \cite{Huege:2016veh}. 
First, the electric field traces simulated in the [$\vec{e}_\mathrm{NS}$, $\vec{e}_\mathrm{WE}$, $\vec{e}_\mathrm{V}$] coordinate system are rotated into the [$\vec{e}_{\vB}$, $\vec{e}_{\vvB}$, $\vec{e}_{\mathrm{v}}$] coordinate system. This allows us to calculate the energy fluence for each of these polarizations \fvB, \fvvB, and \fv, while \fv is almost zero since the electric field of the radio emission is oscillating perpendicular to $\vec{v}$. Then we decompose the signal into one part originating from the geomagnetic \fgeo and another part originating from the charge-excess \fce effects, making use of the known polarization characteristics, i.e., (derived from \cite{Glaser:2018byo})
\begin{gather}
    f_{\mathrm{geo}} = \left(\sqrt{f_{\vec{v}\times\vec{B}}} -  \frac{\cos\phi}{|\sin\phi|} \cdot \sqrt{f_{\vec{v} \times  (\vec{v} \times \vec{B})}} \right)^2 \nonumber \\
    f_{\mathrm{ce}} = \frac{1}{\sin^2\phi} \cdot f_{\vec{v} \times  (\vec{v} \times \vec{B})},
    \label{eq:ldf:posfraction} 
\end{gather}
with $\phi$ the polar angle between an observer position and the positive \vB axis. The underlying concept of Eq.~\eqref{eq:ldf:posfraction} is the following: the strength of the charge-excess emission is solely estimated from the emission in the \vxvxB{-}polarization. Given the position in the shower plane, one can estimate the contribution of the charge-excess emission to the overall emission in the \vB{-}polarization. The disadvantage of this ansatz is obvious: close to or on the \vB axis ($\sin\phi \rightarrow 0$), no signal is polarized in the \vxvxB direction, while the term $1 / \sin^2\phi$ diverges, hence the ansatz loses validity. It should be noted that this ansatz is not affected by the early-late asymmetry as this asymmetry does not affect the polarization of the emission. However, the disentangled signals \fgeo and \fce need to be early-late corrected to show the expected symmetry. A different ansatz that overcomes this problem but comes with other disadvantages is discussed in \cite[Appendix B.1]{Schluter:2022yev} and only mentioned here for completeness. It has to be mentioned that the equations in \eqref{eq:ldf:posfraction} assume that both emissions arrive simultaneously at an observer, i.e., without any phase shift. Such a phase shift would give rise to a circularly polarized component in the incoming electric field which indeed has been seen in experimental data \cite{PhysRevD.94.103010}, i.e., there is a time delay between the pulses originating for the charge-excess and geomagnetic emission. To quantify the fraction of circular polarization in the radio pulses we calculate the Stokes parameters $I$, $Q$, $U$, $V$ following the procedure detailed in reference \cite{PhysRevD.94.103010}. Since the relative strength of the charge-excess emission decreases with the zenith angle (cf. Sec. \ref{sec:ldf:charge_excess_fraction}), the fraction of circularly polarized signal is small for most showers in our set. The determined time delay, following \cite{PhysRevD.94.103010}, is within $\Delta t < \SI{1}{ns}$ for most observers and thus the above equations are applicable for the radio emission in the \SIrange{30}{80}{MHz} band. To ensure that this holds, we only use showers with a geomagnetic angle $\alpha > 20^\circ$, i.e., the angle between the shower axis and the Earth's magnetic field vector, to develop the model.

\begin{figure}[t]
    \includegraphics[width=\textwidth]{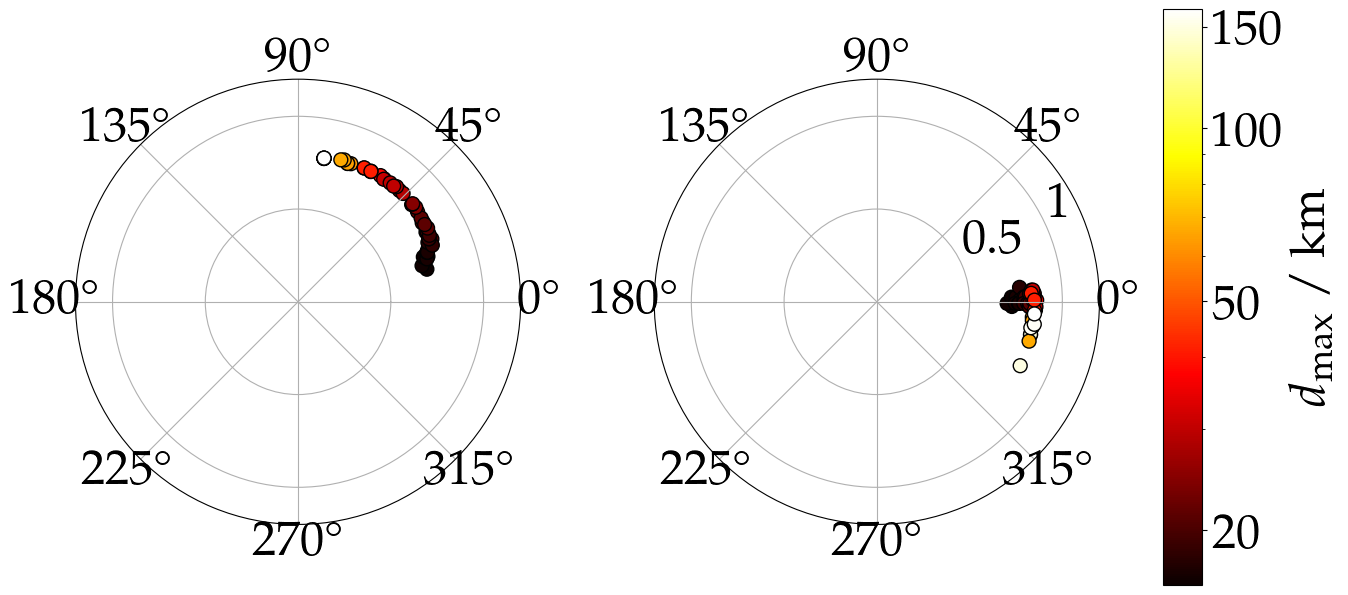}
    \caption{The position of the maximum energy fluence in polar coordinates from air showers arriving from South without (left panel) and with (right panel) early-late correction according to Eq.~\eqref{eq:ldf:early-late}. The coordinates show the rotation for positive \vxB axis (outer axis) and lateral axis distance normalized to $r_0$ according to Eq.~\eqref{eq:ldf:r0} (inner axis). The color code highlights a correlation between the rotation away from the \vxB axis and \dmax.}
    \label{fig:ldf:early-late-maximum}
\end{figure}
\section{Effect of the early-late asymmetry on the emission pattern}
\label{app:ldf:el-angle}
As explained in Sec.~\ref{sec:ldf:emission}, in addition to the asymmetry due to the interference between geomagnetic and charge-excess emission, there is also an early-late asymmetry present in the radio-emission footprints from inclined air showers. The latter disturbs the well-known emission pattern produced by the interference from which the maximum emission is expected at the \vxB axis. In Fig.~\ref{fig:ldf:early-late-maximum} the position of the maximum emission is shown in polar coordinate for uncorrected emission patterns (right panel) and early-late corrected emissions (left panel). In both panels, the outer axis gives the rotation from the \vxB axis in degree, the inner axis gives the lateral axis distance normalized to $r_0$ and the color code shows the \dmax of the respective showers. In the figure, only showers coming from the South are shown as for those the shower axis projected on the ground is in the same plane as the \vxvxB axis and perpendicular to the \vxB axis. As it can be seen, with increased \dmax the maximum is rotated towards the \vxvxB axis, i.e., towards the incoming direction of the air showers. This indicates that the early-late asymmetry is increasingly dominant. If the early-late effects are corrected, the maximum is found around the \vxB axis which is again consistent with the interference pattern from geomagnetic and charge-excess emission only. It is also apparent that for the corrected emission pattern the maximum for showers with a zenith angle of 85$^\circ$ (= $\dmax \sim 150\,$km) is found rotated from the \vxB axis. However, keep in mind that at those inclinations the charge-excess emission is vanishing and hence no clear maximum at the \vxB axis is expected.

\section{Parameterizations for the signal model}
\label{app:ldf:add_param}
In the next section \ref{app:ldf:fgeo_param}, the six parameters of the lateral distribution function \fgs \eqref{eq:ldf:fgs}
describing the shape of the geomagnetic emission are correlated with \dmax. In section \ref{app:ldf:ce-param} the parameterization of the charge-excess fraction $a_\mathrm{ce}^\mathrm{ICRC19}$ \eqref{eq:aparam} is optimized to obtain the new expression $a_\mathrm{ce}$ \eqref{eq:ldf:charge-excess-param}.

\subsection{Parameterizations of the shape of the geomagnetic emission}
\label{app:ldf:fgeo_param}
\begin{figure}[t]
    \centering
    \includegraphics[width=0.85\textwidth]{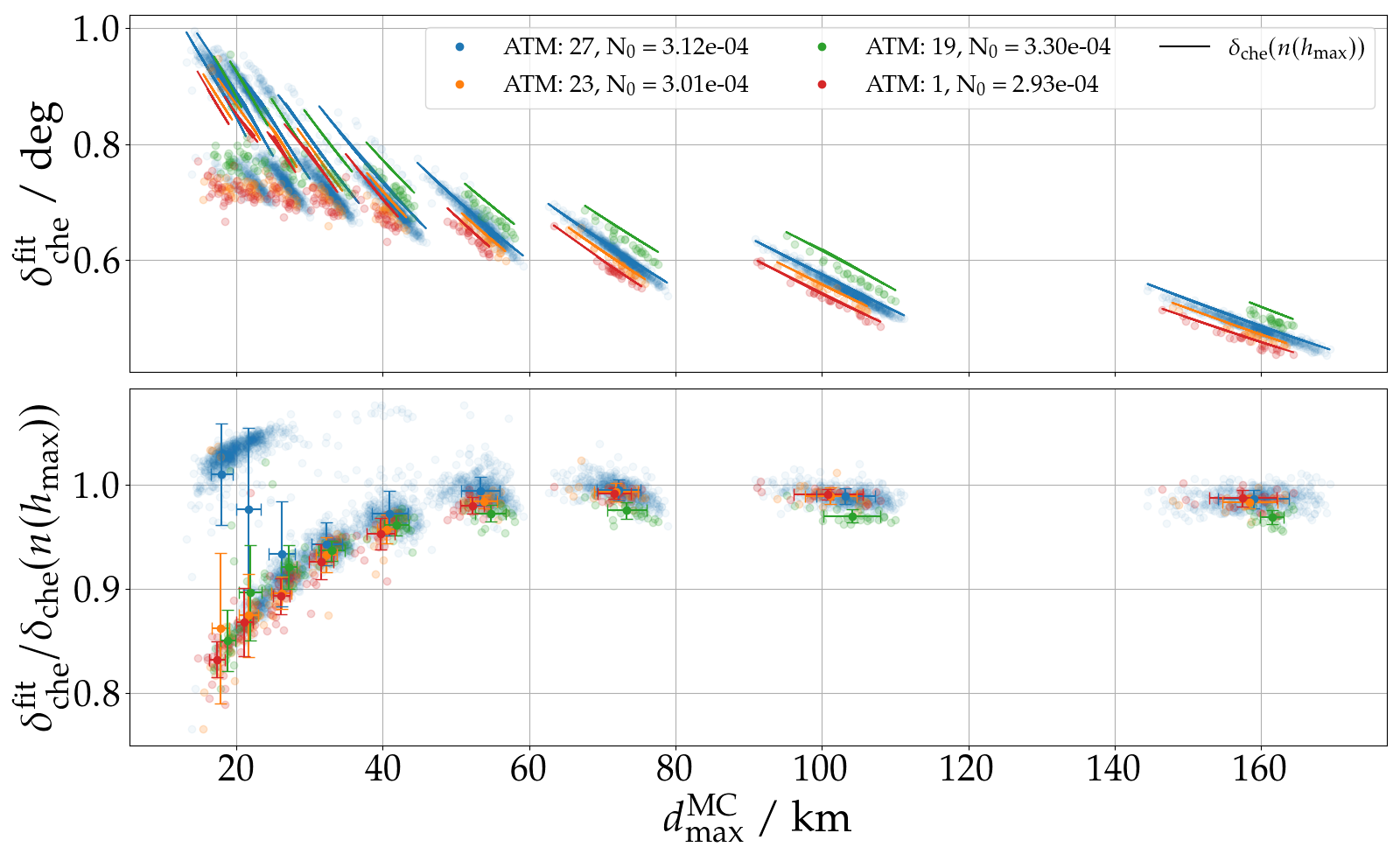}
    \caption{The top panel shows $r_0^\mathrm{fit}$ converted into an opening angle $\delta_\mathrm{Che}^\mathrm{fit}$ using \dmax (dots) compared to the Cherenkov angle $\delta_\mathrm{Che}$ as calculated from the refractive index $n$ at shower maximum by Eq.~\eqref{eq:ldf:r0} (lines) for four different atmospheres US standard: 1, Auger February: 19, Auger June: 23, Auger October: 27 (the refractivity at sea level $N_0 = n_0 - 1$ used in combination to the density profiles are quoted in the legend). The bottom panel shows the deviation between fitted and calculated values. The squared markers show the mean and standard deviations of the deviation.}
    \label{fig:ldf:r0}
\end{figure}

First, we investigate how the radius of the Gaussian $r_0^\mathrm{fit}$ relates to \dmax. In Fig.~\ref{fig:ldf:r0}, the opening angle of a cone originating at \dmax, $\delta^\mathrm{fit}_\mathrm{Che}(r_0^\mathrm{fit}) = \tan(r_0^\mathrm{fit} / \dmax)$, calculated from the fitted radius $r_0^\mathrm{fit}$, is shown (top panel) as a function of \dmax (dots). The prediction for the Cherenkov angle $\delta_\mathrm{Che}$ according to Eq.~\eqref{eq:ldf:r0} is shown for comparison (lines). Both, $\delta^\mathrm{fit}_\mathrm{Che}$ and the theoretical prediction are shown for four different simulated atmospheres. The atmospheres at the location of the Pierre Auger Observatory for February (summer), July (winter), and October correspond to the maximum, minimum, and yearly average for the refractivity at ground level, respectively \cite{PhDGlaser}. The bottom panel shows the relative deviation between fitted and predicted angles. The comparison shows an overall remarkable agreement for larger zenith angles and different atmospheres. For lower zenith angles, a systematic deviation can be found. However, it is possible to use $r_0$ determined as a function of \dmax according to Eq.~\eqref{eq:ldf:r0} instead of fitting it without losing significant accuracy. We carefully checked that the remaining free parameters sufficiently compensate for the deviations introduced when using the predicted value of $r_0$. In the following, we refer to $r_0$ as the Cherenkov radius.
\begin{figure}
    \centering
    \includegraphics[width=0.49\textwidth]{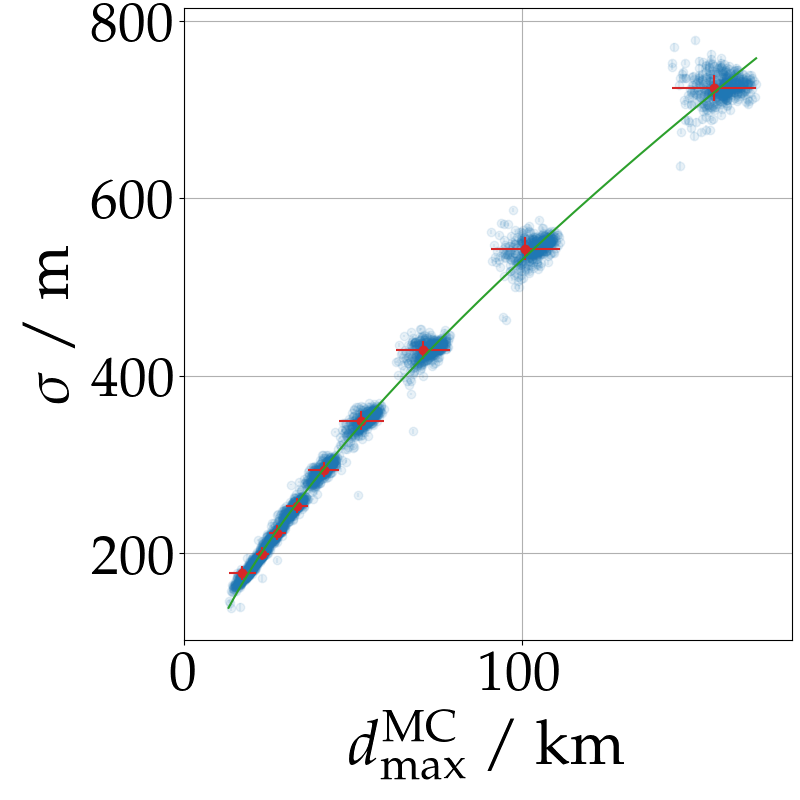}\hfill
    \includegraphics[width=0.49\textwidth]{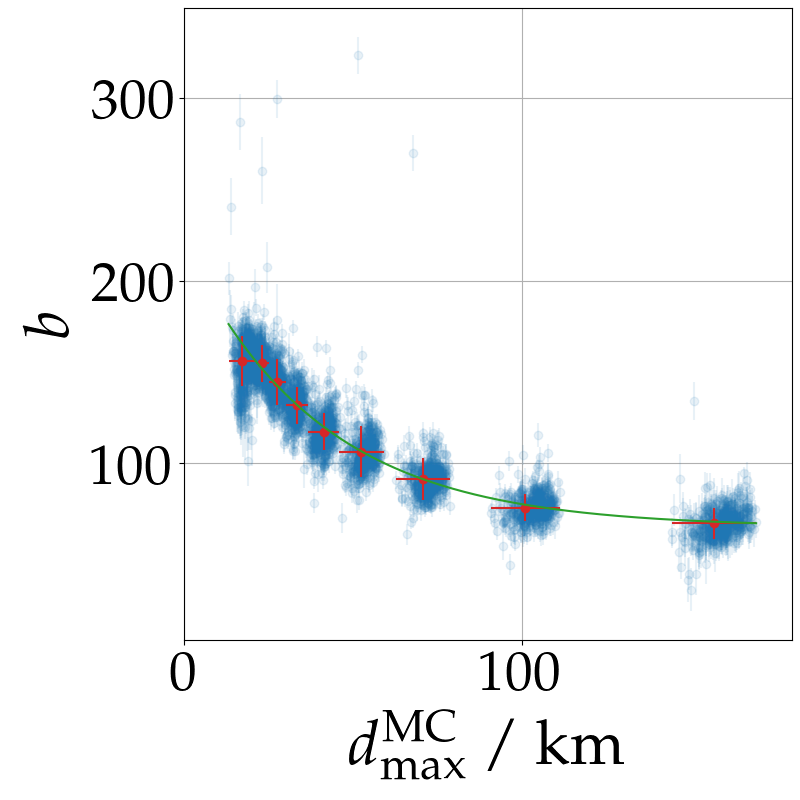}\\
    \includegraphics[width=0.49\textwidth]{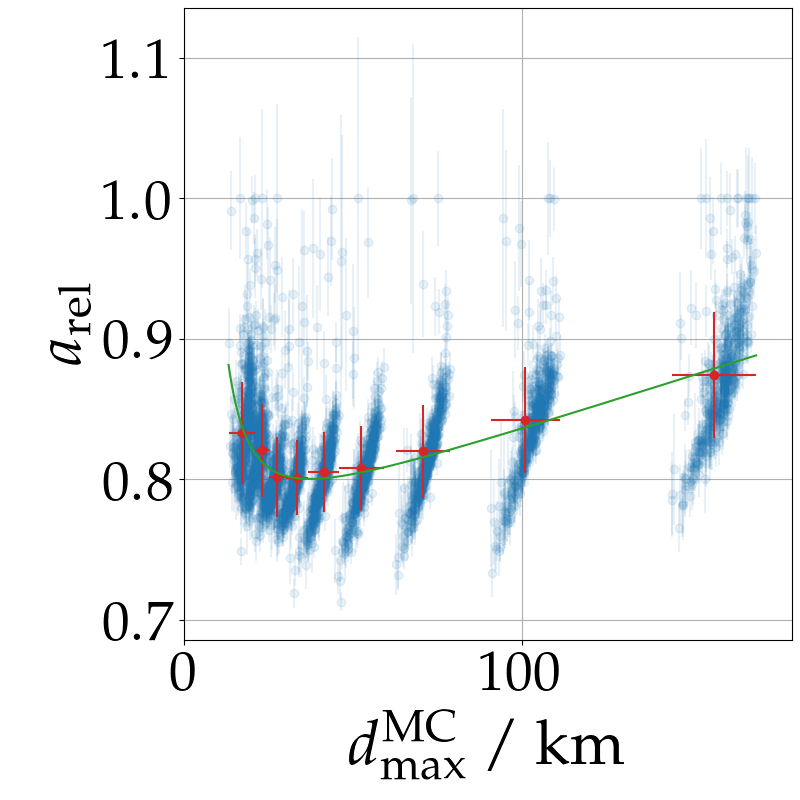}\hfill
    \includegraphics[width=0.49\textwidth]{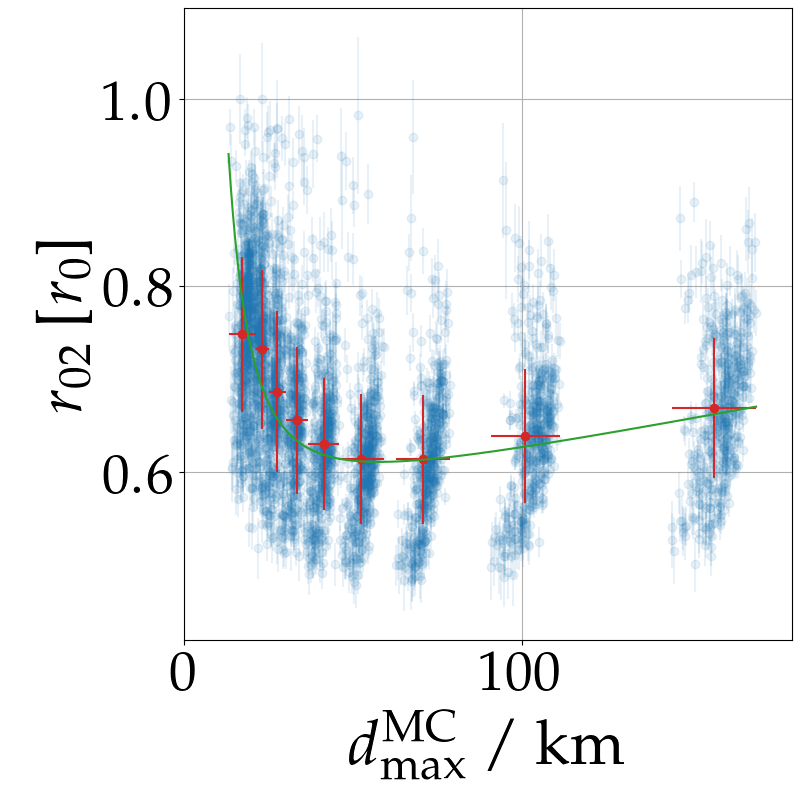}
    \caption{Parameterizations according to Eqs.~\eqref{eq:ldf:sigma} to \eqref{eq:ldf:r02} (green lines) compared with the fit values for individual simulations (blue points) as well as their profiles (red points; means and standard deviations).}
    \label{fig:ldf:parameterizations}
\end{figure}

Next, we study the correlation of $\sigma$ in Eq.~\eqref{eq:ldf:fgs} with \dmax. The top-left panel in Fig.~\ref{fig:ldf:parameterizations} shows the values derived for $\sigma(\dmax)$ when fitting all showers with $r_0$ fixed to Eq.~\eqref{eq:ldf:r0} and the slope $s = 5$ (blue circles). The red markers illustrate the mean and standard deviation (vertical error-bars, the bin sizes are indicated by the horizontal error-bars) of the fitted data. The green line shows our parameterization for $\sigma$, 
\begin{equation}
    \sigma = \left(0.132 \cdot \left(\dmax - 5\,000\,\mathrm{m}\right) ^ {0.714} +  56.3\right)\,\mathrm{m}.
    \label{eq:ldf:sigma}
\end{equation}
We normalize the function with the term ``$\dmax - 5\,000\,\mathrm{m}$'' to decrease the statistical fluctuations in the fitted parameters. However, this restricts the parameterization to values of \dmax > 5$\,000\,$m. While \dmax < 5$\,000\,$m is very unlikely for hadron-initiated air showers with zenith angles $\theta \geq 60^\circ$ as it would require depths of \Xmax > \SI{1200}{\gcm}, for neutral particles, in particular neutrinos, \dmax < 5$\,$km is not difficult to imagine. However, if we assume that the radio emission is still detectable at a 4$^\circ$ off axis angle, the maximum axis distance for $\dmax = 5\,$km is $\sim 350\,$m, which is too small to detect such showers in more than one or two antennas with a kilometer-spaced detector. The uncertainties of the fitted data are statistical, estimated from the $\chi^2$-minimization of the LDF fit. They can not explain the deviation of single data points from the parameterization. It can not be excluded that those points represent an alternative minimum. However, the global minimum can be easily identified with the parameterization by the vast majority of the data. To obtain the optimal values for the parameterizations in Eq.~\eqref{eq:ldf:sigma} we employed again a $\chi^2$-minimizations, this time using the \textit{iminuit} python package \cite{iminuit}.

The same procedure is now applied consecutively to the parameters $p(r)$ (resp. $b$), $a_\mathrm{rel}$, and $r_{02}$, in this order. Their distributions are shown in Fig.~\ref{fig:ldf:parameterizations} and their parameterizations are given by Eqs.~\eqref{eq:ldf:p} - \eqref{eq:ldf:r02}:
\begin{align}
    p(r) = \left\{ \begin{array}{cc}
        2                              & \hspace{5mm} r \leq r_0 \\
        2 \cdot (r_0 / r) ^ {b / 1000} & \hspace{5mm} r > r_0
    \end{array} \right. \text{,} \;\; b = 154.9 \cdot \exp\left(-\frac{\dmax}{40.0\,\mathrm{km}}\right ) + 64.9,
    \label{eq:ldf:p}
\end{align}

\begin{equation}
    a_\mathrm{rel} = 0.757 + \frac{\dmax}{1301.4\,\mathrm{km}} + \frac{19.8\, \mathrm{km}^2}{\dmax^2},
    \label{eq:ldf:arel}
\end{equation}

\begin{equation}
    r_{02} = 0.552 + \frac{\dmax}{1454.2\,\mathrm{km}} + \frac{66.2\, \mathrm{km}^2}{\dmax^2}. \label{eq:ldf:r02}
\end{equation}
In the distributions for $a_\mathrm{rel}$ and $r_{02}$, an additional trend, not described by the parameterizations, is significant. Within one zenith angle bin, a steep increase of the corresponding parameter from deep to shallow showers is apparent. The matter is further discussed in the last paragraph of this section, for now we choose to only describe the correlation of all parameters with \dmax.

\begin{figure}
    \centering
    \includegraphics[width=\textwidth]{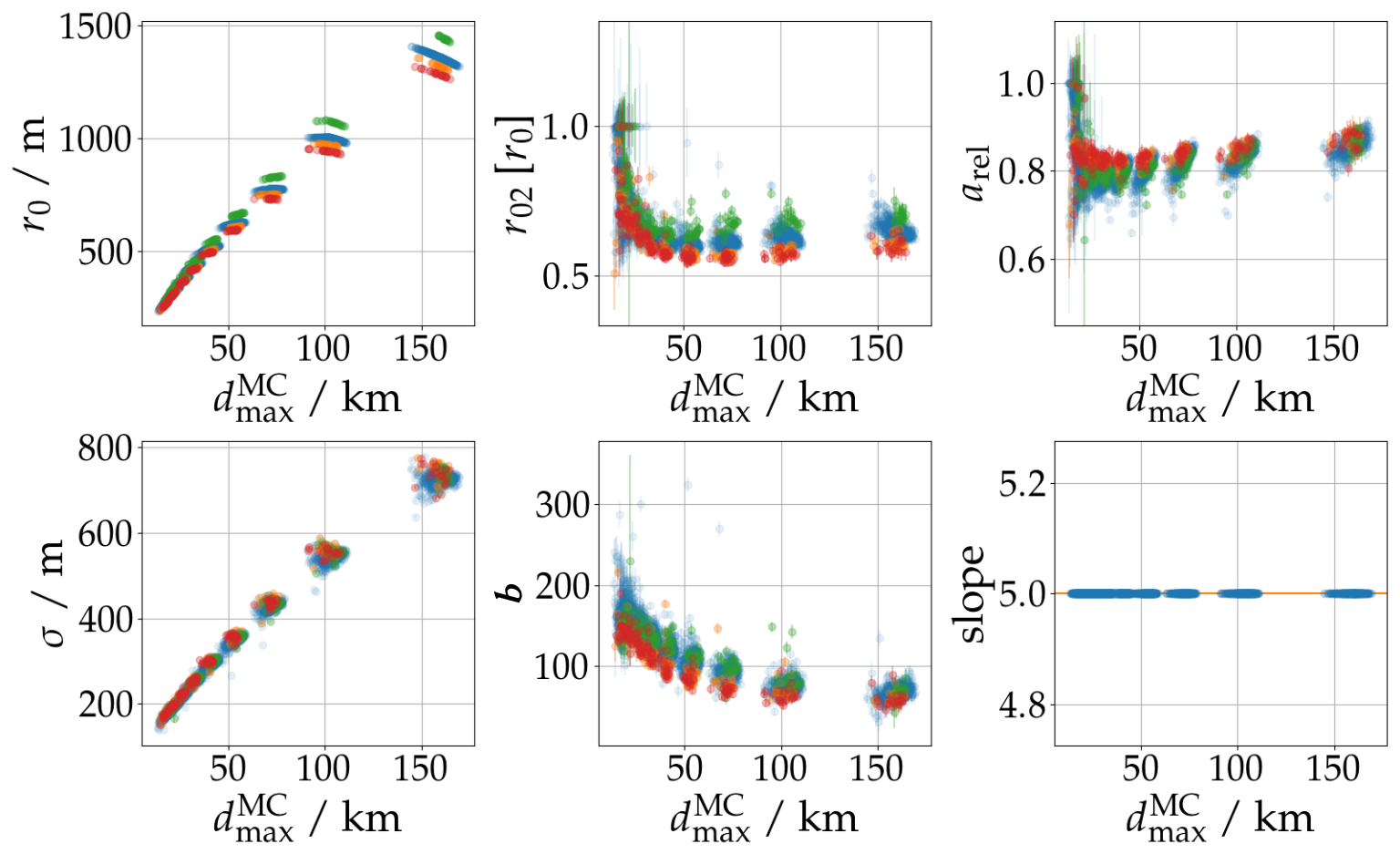}
    \caption{Parameters of \fgs Eq.~\eqref{eq:ldf:fgs} extracted from simulations with different atmospheric conditions (same color mapping as in Fig.~\ref{fig:ldf:r0}). The distributions correspond to fits where only $s=5$ and $r_0$ according to Eq.~\eqref{eq:ldf:r0} are constrained.}
    \label{fig:ldf:parameterizations_atm}
\end{figure}

We also verified the fit results for different atmospheres. With the prediction of $r_0$ depending on the atmospheric profile, the parameterization of the LDF \fgs explicitly uses information of the atmosphere. The other parameters, however, are assumed to be universal, i.e., do not depend (significantly) on the simulated atmosphere. In Fig.~\ref{fig:ldf:parameterizations_atm}, the correlation of the parameters with \dmax for the different simulated atmospheres is shown. Although the atmosphere influences the correlations of the parameters with \dmax, the variation is minimal and the October atmosphere used for the parameterization indeed describes the mean reasonably well. The effect on the geomagnetic radiation energy was found to be below 1\%. However, it should be kept in mind, that this conclusion was obtained with star-shaped simulation which for this purpose has limited informative value.

The observed \Xmax-dependent bias in the energy reconstruction can be resolved, to a large degree, by describing the secondary correlation of \Xmax with the LDF parameters $a_\mathrm{rel}$ or $r_{02}$, cf.\ Fig.~\ref{fig:ldf:parameterizations}. The secondary correlation can be explained by the ambiguity of \dmax for different zenith angles and \Xmax values. This ambiguity, although not completely resolved, can be reduced using the density at the shower maximum \rhomax as observable. An elegant solution to resolve the ambiguity is the introduction of $d_{750}$ the distance between shower core and a fixed slant depth of \SI{750}{\gcm}, in the parameterizations of $a_\mathrm{rel}$ or $r_{02}$. However, when doing so, we found an implausible kink in the distribution of the fitted \dmax distribution with simulations from the validation set (cf.\ Fig.~\ref{fig:ldf:density}, \textit{right}), and for this reason, the fact that we do not observe a significant primary-particle dependent bias and for the sake of simplicity we decided to not include such a term in the parameterizations. However, if needed in the future, our model can be improved by a more thorough study of these secondary correlations.

\subsection{Refined parameterization of the charge-excess fraction}
\label{app:ldf:ce-param}
\begin{figure}[t]
    \centering

    \includegraphics[width=0.48\textwidth]{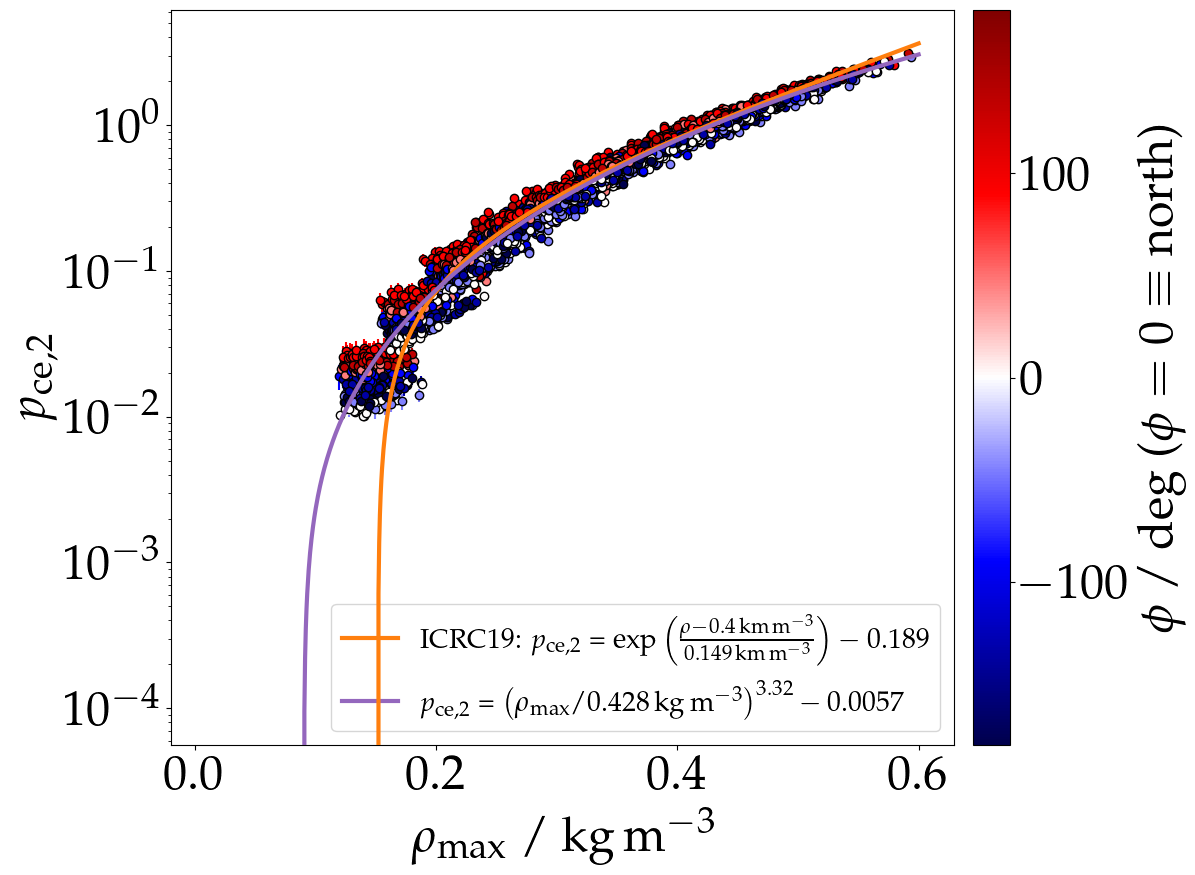}\hfill
    \includegraphics[width=0.48\textwidth]{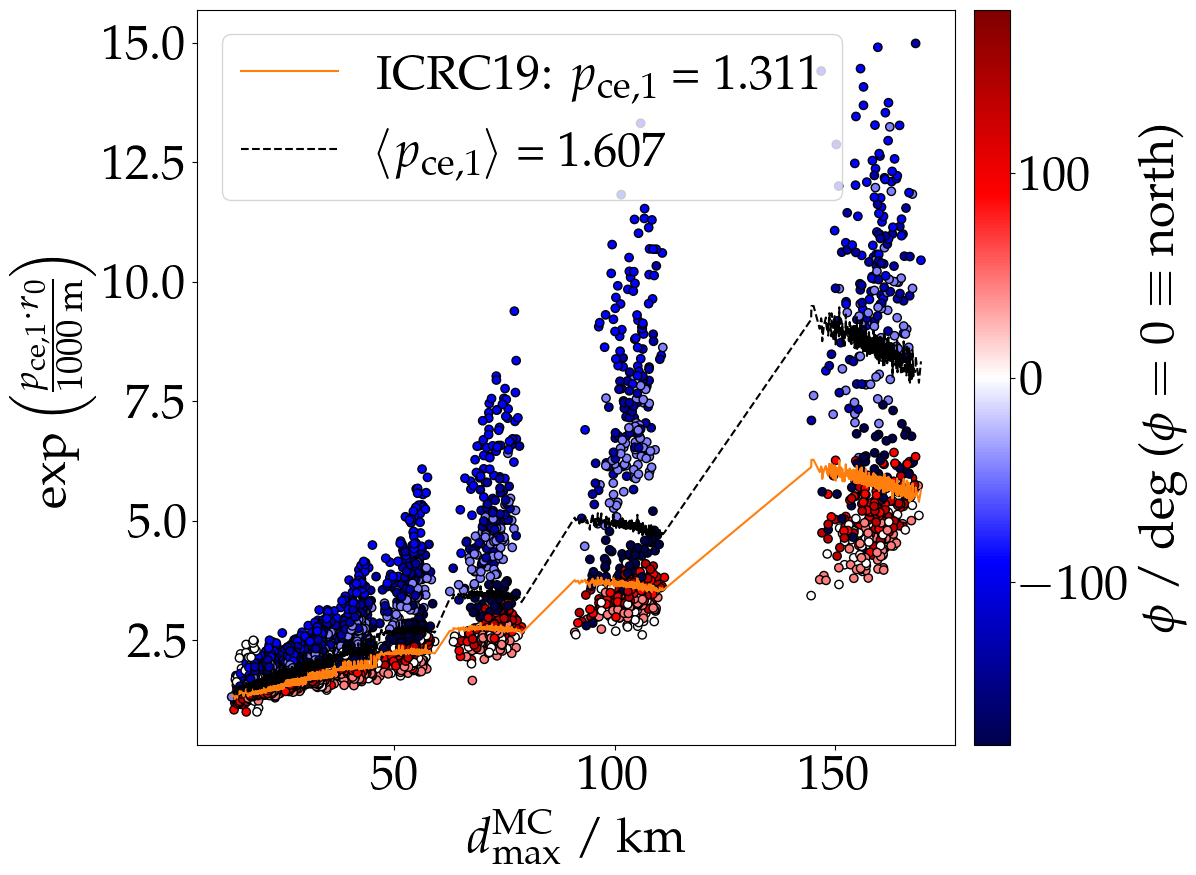}\\
    \includegraphics[width=0.48\textwidth]{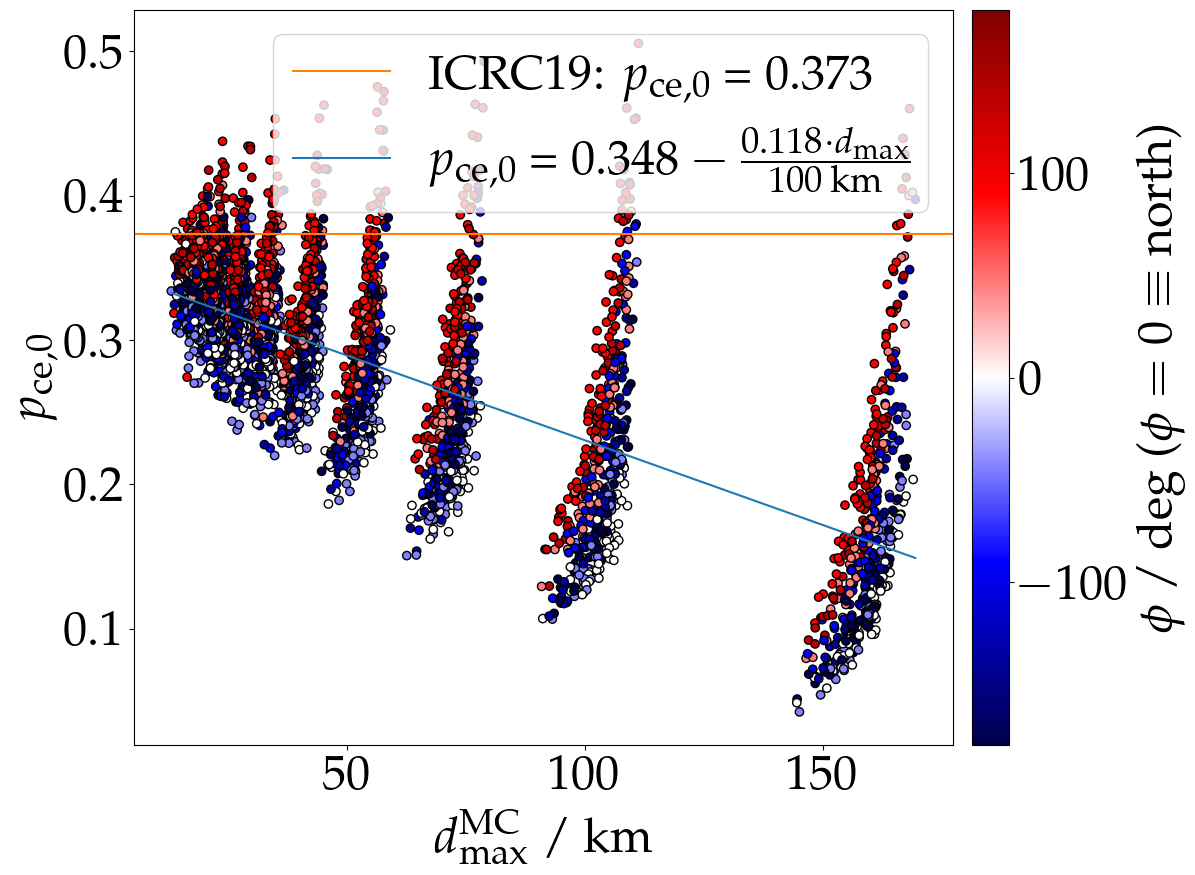}\hfill
    \includegraphics[width=0.4\textwidth]{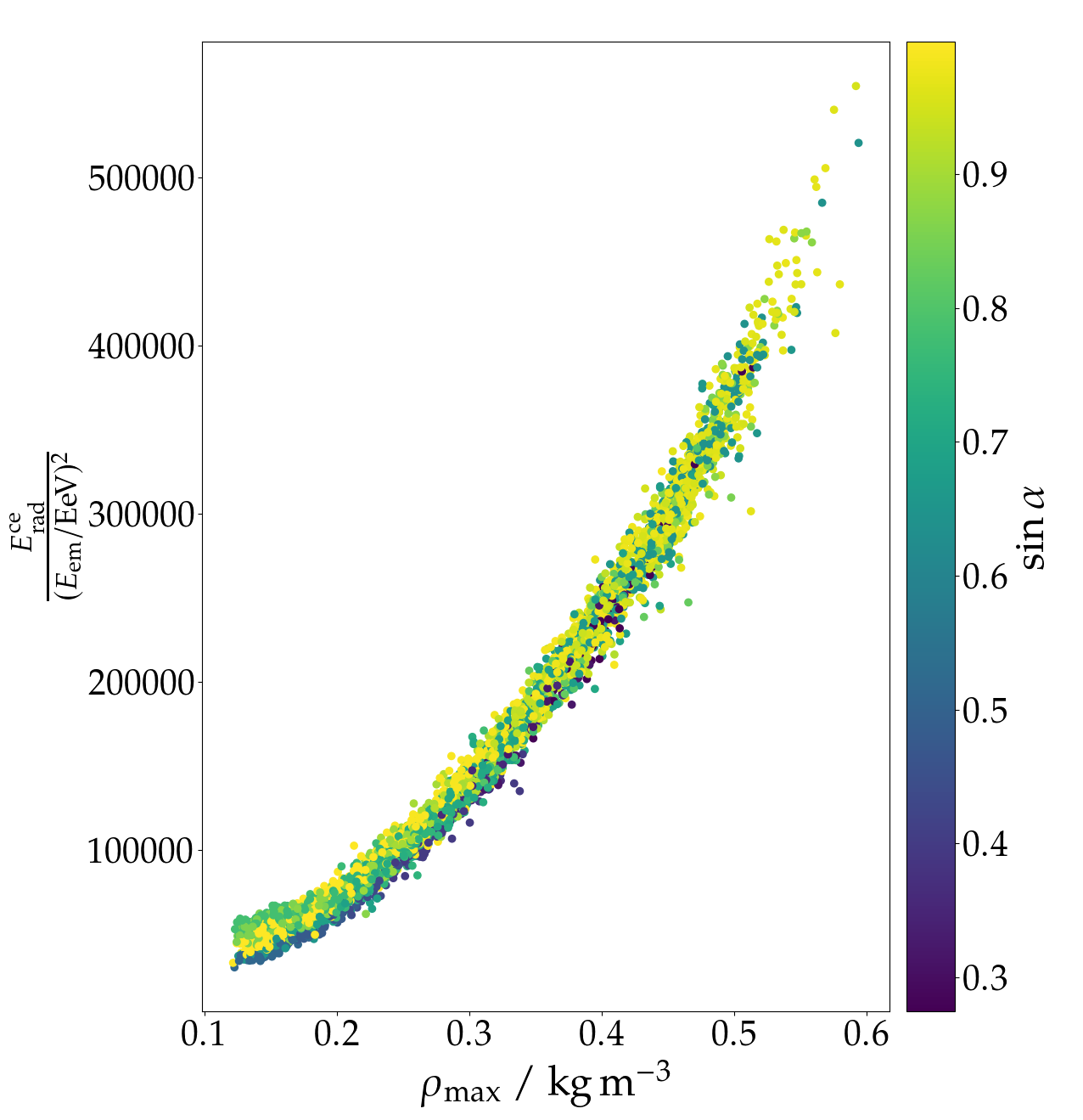}\hspace{1cm}

    \caption{\textit{Top-left}: Optimized density scaling of the charge-excess fraction as function of the air density at the shower maximum \rhomax. The color code denotes the shower arrival direction. Lines are explained in the text. \textit{Top-right}: Optimized parameter describing the exponential correction for large axis distances. \textit{Bottom-left}: Optimized parameter describing the linear scaling of the charge-excess fraction with the of axis angle. \textit{Bottom-right}: Total charge-excess radiation energy as function of the air density at the shower maximum \rhomax. A strong correlation is visible, the radiation decreases with decreasing density, i.e., increasing zenith angle. The color code highlighting the sine of the geomagnetic angle shows no correlation (as expected).}
    \label{fig:ldf:charge_excess_fraction_param}
\end{figure}

To refine the parameterization of the charge-excess fraction $a_\mathrm{ce}^\mathrm{ICRC19}$ \eqref{eq:aparam}, we optimize the different terms $p_{\mathrm{ce}, i}$ in the parameterization by fitting the distribution of symmetrized signals, i.e., signals for which the charge-excess emission has been subtracted using the parameterization, to the rotational symmetric, fully-parameterized LDF from Sec.~\ref{sec:ldf:geomagnetic_emission}. First, we optimize the term describing the scaling of the charge-excess fraction with the air density $p_{\mathrm{ce}, 2}$. While optimizing Eq.~\ref{eq:ldf:f_geo}, i.e., $\fgeo(\ace(p_{\mathrm{ce}, 2}))$, to minimize the difference to the constrained \fgs, we find the value for $p_{\mathrm{ce}, 2}$ for which the estimated geomagnetic emission is the most symmetric. Only the normalization $f_0$ of \fgs is varied as well. Fig.~\ref{fig:ldf:charge_excess_fraction_param} (\textit{Top-left}) shows the correlation of $p_{\mathrm{ce}, 2}$ obtained for all showers with \rhomax. The purple curve shows our new description given by
\begin{equation}
    \label{eq:ldf:charge-excess-fraction-density}
    p_{\mathrm{ce}, 2} = \left(\frac{\rhomax}{0.428\,\kgm}\right)^{3.32} - 0.0057.
\end{equation}
The functional form is rather ad-hoc but describes the data better than the exponential function used at the ICRC19 \cite{Huege:2019cmk} which is shown by the orange curve. Also, this new function can become negative at small \rhomax, and thus implausible, but does so for lower values $\rhomax < 0.09\,\kgm$ than the exponential model used in the ICRC19-parameterization which becomes negative for $\rhomax \lesssim 0.15\,\kgm$. This allows us to extend the parameterization to zenith angles of \SI{85}{\degree} and beyond.
Similarly, the ``exponential correction'' $p_{\mathrm{ce}, 1}$ term and the ``off-axis angle'' $p_{\mathrm{ce}, 0}$ term are substituted and refined (in this order). For $p_{\mathrm{ce}, 1}$ the mean is used. In Fig.~\ref{fig:ldf:charge_excess_fraction_param} (\textit{Top-right}) the total exponential term is shown using $r=r_0$ per shower and compared to the previous value. For the off-axis angle term, instead of a constant factor a linear model with a slope depending on \dmax is used now, cf.\ Fig.~\ref{fig:ldf:charge_excess_fraction_param} (\textit{Bottom-left}). 

\section{Density scaling of the geomagnetic and charge-excess emission}
\label{app:ldf:emission-density-scaling}
While the air-density scaling of the overall emitted radiation energy has been investigated earlier \cite{Glaser:2016qso} and found to coincide with the scaling of the dominant geomagnetic emission, the correlation of the charge-excess emission with the air density has not previously been studied that thoroughly. We extract the charge-excess emission by Eq.~\eqref{eq:ldf:posfraction} from star-shaped simulations and perform a 2d-spatial integration over the interpolated footprint to estimate its radiation energy. We interpolate the 2d-footprint via Fourier decomposition \cite{Fourier}. We find that the charge-excess emission decreases in absolute strength (and not only relative strength) with increasing zenith angle, i.e., decreasing density at the shower maximum, see Fig.~\ref{fig:ldf:charge_excess_fraction_param} (\textit{bottom-right}). A similar correlation is reported in \cite[Fig.\ 2]{Chiche:2021iin}. An explanation for this phenomenon might be that in a denser atmosphere, i.e., for more vertical showers, more electrons are ionized from the ambient atmosphere hence the negative charge excess is stronger. However, this simple explanation needs verification.

The scaling of the geomagnetic emission with the density of the shower maximum has already been shown in Fig.~\ref{fig:ldf:density} (\textit{left}). It can be explained with the following picture: The emission strength depends on the mean free path length with which the electromagnetic particles traverse the atmosphere. With a larger mean free path, equivalent to traversing a less dense atmosphere negatively and positively charged particles can drift further apart before interacting, creating a stronger transverse current and thus resulting in a stronger geomagnetic emission. For a given slant depth, the density at the shower maximum \rhomax, is smaller for larger zenith angles and larger \dmax, respectively.


\section{Reconstructing the electromagnetic shower energy}
\subsection{Deriving the true electromagnetic shower energy from CORSIKA simulations}
\label{app:ldf:energy}
The strength of the radio emission is strongly correlated with the energy of the electromagnetic shower cascade, i.e., the electromagnetic shower energy. It should be stressed that this is slightly different from the fluorescence light seen by optical telescopes which better correlates with the total calorimetric energy (for which other particles like muons have a non-negligible contribution). We compute the electromagnetic energy as the sum over the longitudinal energy deposit $E_i$ for gamma rays, electrons, and positrons (ionization and cut) as provided in the CORSIKA \texttt{DATnnnnnn.long} files, i.e.,
\begin{equation}
    \label{eq:ldf:eem_mc}
    E_\mathrm{em}^\mathrm{MC} = \sum_{i=0}^N E_i(\gamma) + E_i^\mathrm{ioniz.}(e^+e^-) + E_i^\mathrm{cut}(e^+e^-).
\end{equation}
It is worth noting that this includes the energy deposit in the ground plane (which is accounted for in the two last rows of this table with the SLANT option). In inclined air showers, no clipping effects of the radio emission occur,  because the showers can evolve fully before the ground is reached.

\subsection{Reconstruction of the electromagnetic shower energy for showers generated with Sibyll2.3d}

\begin{figure}
    \centering
    \includegraphics[width=\textwidth]{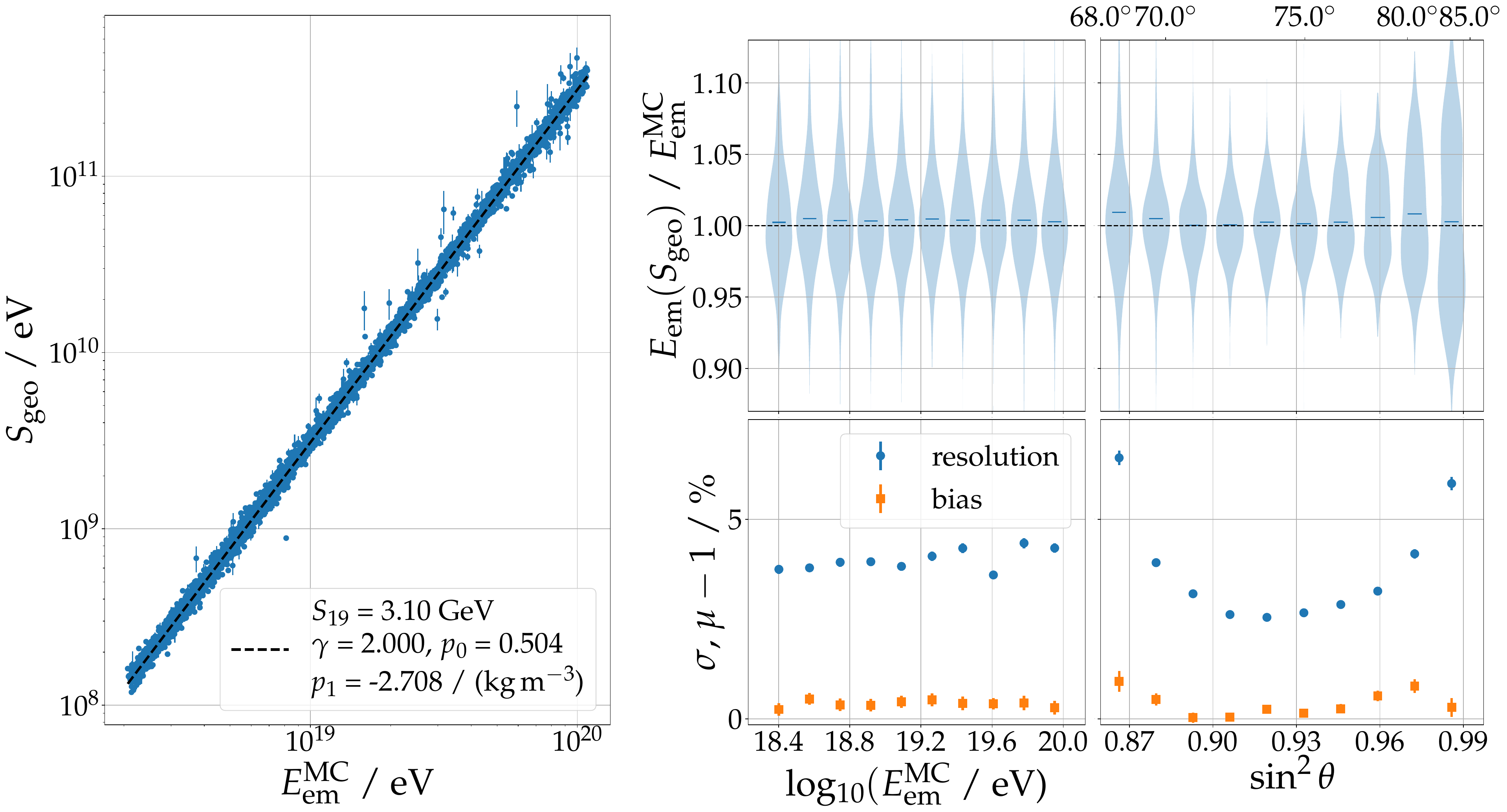}
    \caption{Reconstruction of the electromagnetic shower energy \Eem for the showers generated with Sibyll2.3d. See caption of Fig.~\ref{fig:ldf:energy}.}
    \label{fig:ldf:sibyll}
\end{figure}

Fig.~\ref{fig:ldf:sibyll} shows the \Eem reconstruction for showers generated with the Sibyll2.3d high-energy interaction model. From the 6199 showers with a zenith angle greater than \SI{68}{\degree} and at least 5 simulated observers, 6185 showers were reconstructed with good quality. The results are very comparable to the ones archived with the QGSJETII-04 showers: The $S_{19}$ parameter decreased slightly by less than 2\% and the resolution at lower zenith angles worsens slightly. It is worth stressing that the simulations used to develop this model were solely generated with QGSJETII-04, hence a small decrease in reconstruction quality for Sibyll-generated showers is not surprising. Nevertheless, this result underlines the fact that the radio emission has little dependence on the underlying hadronic interaction model as long as one normalizes quantities to the electromagnetic energy in the air shower.
\label{app:ldf:energy-sibyll}

\bibliographystyle{unsrt}
\bibliography{references}

\end{document}